\documentstyle[12pt,epsf,preprint,aps]{revtex}

\begin{document}

\newcommand\beq{\begin{equation}}
\newcommand\eeq{\end{equation}}

\textwidth=16.0cm
\textheight=23.5cm 

\draft 

\title{
Protoneutron Stars with Kaon Condensation\\
and their Delayed Collapse }
\author{Masatomi Yasuhira and Toshitaka Tatsumi}
\address{Department of Physics, Kyoto University,\\
Kitashirakawa-Oiwake-cho,\\
Kyoto 606-8502, Kyoto, JAPAN\\
e-mail:{\it yasuhira@ruby.scphys.kyoto-u.ac.jp}}

\maketitle

\begin{abstract}
Properties of protoneutron stars are discussed
in the context of kaon condensation.
Thermal and neutrino-trapping effects are very
important ingredients to study them.
By solving the TOV equation,
we discuss the static properties of protoneutron stars
and the possibility of the delayed collapse
during their evolution.


\end{abstract}
\newpage

\section{Introduction}

A protoneutron star (PNS),
which is formed after the gravitational collapse
of massive stars,
consists of hot and dense hadronic matter.
At birth
the matter is lepton rich 
due to the trapping of neutrinos
and has entropy per baryon of order 1-3
in units of Boltzmann constant.
A PNS evolves in a few tens of seconds 
through the deleptonization and initial cooling eras
to usual cold neutron stars\cite{Prakash}.
During the evolution,
some of the stars may become to be gravitationally unstable
due to some hadronic phase transitions
and collapse to the low-mass black holes\cite{BrownBethe},
which we call the delayed collapse.

To make the low-mass black hole
with the gravitational mass of $O(1.4M_\odot)$,
first, the maximum mass of a PNS
must be sufficiently low, and secondly,
the mass must exceed the maximum-mass during their
evolution due to neutrino emission,
cooling and/or fall-back of matter.
The possibility of making the low-mass black holes
in supernovae and its relation with SN1987A
has been discussed by many authors\cite{WoosZampFry}.
The observation of neutrinos from SN1987A
suggested a formation of a PNS in it.
However the neutron star has never been detected so far.
One idea to resolve this mystery is that
the neutron star must have collapsed further
into a low-mass black hole at some later time
because of either the accretion of a sufficient
fall-back mass and/or some change of its equation of state(EOS)

Thus it is very important to study the change of EOS
during the evolution of a PNS
to discuss the possibility of the delayed collapse:
the EOS depends on what kind of hadronic phase transitions
takes place:
quark matter, boson condensed matter, hyperonic matter
and so on\cite{Prakash}.
There neutrino-trapping and thermal effects
play important roles
in the rearrangement of constituents of matter
and they are also crucial for
whether phase transition occurs or not.
Numerical simulations
to describe the evolution of a PNS
and the delayed collapse
have been done\cite{Baumgarte}\cite{Keil}\cite{PonsH}\cite{PonsKsim}.
In this paper we study kaon condensation
in hot and neutrino-trapping matter,
which may be a key object to the delayed collapse
due to the large softening of
the EOS\cite{BrownBethe}\cite{MYTTC}.

Kaon condensation
has been studied by many authors
mainly at zero temperature\cite{CHLee},
since first suggested by Kaplan and Nelson\cite{Kaplan}.
Its implications have been also discussed mainly
in the context of cold neutron stars;
maximum-mass, cooling mechanism,
rapidly rotating pulsars and so on.
Baumgarte et al.\cite{Baumgarte}
performed a numerical simulation
of the evolution of a PNS by taking into account
the change of EOS due to kaon condensation
and examined how they collapse to black holes.
However they used the EOS for kaon condensed matter
at {\it zero} temperature.
There has been no study to treat the thermal effect
consistently in the chiral model.

Recently we have presented
a new framework to treat
thermal and quantum fluctuations around the condensate
on the basis of chiral symmetry
\cite{TTMY98l}\cite{procI}\cite{TTMY}.
Making full use of the group structure of chiral manifold,
we have derived the thermodynamic potential
from the nonlinear chiral Lagrangian.
The Goldstone mode arises from the V-spin symmetry
breaking there.
We have found the dispersion relation for this mode,
and that it plays an important role
at finite temperature\cite{TTMYC}.
Here we study the kaon condensation at finite temperature
using the chiral model
and discuss the properties of the PNS.

There are many studies about the kaon condensation at finite
temperature by using the meson-exchange model\cite{Prakash}\cite{Ponsprep}.
Earlier work suggested that the thermal effect on the EOS
is relatively small compared with the neutrino-trapping effect\cite{Prakash}.
Pons et al. also
discussed the EOS for kaon condensed matter
and properties of the PNS\cite{Ponsprep}.
They concluded the thermal effect
is a key object to induce the delayed collapse.
The most important difference between the meson-exchange model
and the chiral model we use here
is about the nonlinearity of the kaon field.
Since the chiral symmetry dictates
the nonlinear Lagrangian for the kaon field,
the properties of the well-developed kaon condensed phase
become very different between both models,
while those are almost the same as far as the
condensate is week\cite{Prakash}\cite{Ponsprep}.
As we shall see in this paper
the chiral model gives a rather stronger condensation
in comparison with the meson-exchange model.
Thereby the chiral model may give a qualitatively different
scenario for the delayed collapse of the PNS.

It is found that
a normal PNS without any phase transition
cannot collapse under no accretion
and the delayed collapse is possible
by the neutrino-trapping effect in the deleptonization era
due to the occurrence of kaon condensation
using a most reasonable parameter for the $KN$ sigma term.
We find that the neutrino-trapping effect
gives a main contribution to the delayed collapse
rather than the thermal effect,
within the chiral model.
We also find, with different parameters,
that the thermal effect may be important
as well as the neutrino-trapping effect.

In Sect.\ref{formulation} we briefly review the
essential part of the formulation
that enables us to treat thermal and quantum fluctuations
around the condensate transparently.

In Sect.\ref{sect:nume} we discuss some features
of EOS in the chiral model and the properties of a PNS,
especially the possibility of its delayed collapse.

Summary and concluding remarks
are given
in Sect.\ref{sect:sum}.

\section{Formulation}\label{formulation}

Recently we have presented a framework to take into account the
quantum and/or thermal fluctuations around the condensate,
in accordance with chiral symmetry \cite{TTMY98l}\cite{procI}\cite{TTMY}. 
The effective partition function
$Z_{chiral}$ at temperature $T$ can be written as
\beq
  Z_{chiral}=N\int [dU][dB][d\bar B] \exp\left[\int_0^\beta d\tau\int d^3x
   \left\{{\cal L}_{chiral}(U, B)+\delta{\cal L}(U, B)\right\}\right],
  \label{eq:part}
\eeq
with the imaginary time $\tau=it$
and
$\beta=1/T$. 
${\cal L}_{chiral}(U, B)={\cal L}_0(U, B)+{\cal
L}_{SB}(U, B)$ is the Kaplan Nelson Lagrangian\cite{Kaplan}
and is represented with 
the octet baryon field, $B$, and the chiral field, 
$U=\exp[2iT_a\phi_a/f]\in SU(3)$,  with $SU(3)$ generators $T_a$ and 
the Goldstone fields $\phi_a$. $\delta{\cal L}$ is the induced SB term as
a result of introduction of chemical potentials,
$\mu_K$ and $\mu_n$,
\begin{eqnarray}
 \delta {\cal L}
  &=& 
  -\frac{f^2\mu_K}{4}{\rm tr}\{[T_{em}, U]
  \frac{\partial U^\dagger}{\partial\tau}+
  \frac{\partial U}{\partial\tau}[T_{em}, U^\dagger]\}
\nonumber\\
 &&{} -\frac{\mu_K}{2}{\rm tr}
  \{B^\dagger[(\xi^\dagger[T_{em}, \xi]+\xi[T_{em}, \xi^\dagger]), B]\}
\nonumber\\
 &&{}-\frac{f^2\mu_K^2}{4}{\rm tr}\{[T_{em}, U][T_{em}, U^\dagger]\}
  +\mu_n{\rm tr}\{B^\dagger B\}-\mu_K{\rm tr}\{B^\dagger[T_{em},B]\} 
\label{del}
\end{eqnarray}
with $U=\xi^2$
and $T_{em}=diag(2/3, -1/3, -1/3)$.
Generally it is complicated to perform the path integral
by separating the fields $\phi_a$
into the sum of classical fields $\langle \phi_a \rangle$
and fluctuation fields $\tilde\phi_a$
in the standard way, because $\{ \phi_a \}$ reside on the 
curved manifold $SU(3)\times SU(3)/SU(3)$.
Instead, we introduce the
local coordinates around the condensed point
to eliminate the curvature effect
in the neighborhood,
which is equivalent to the
following parameterization for $U$;
\beq
  U=\zeta U_f\zeta(\xi=\zeta U_f^{1/2} u^\dagger=u U_f^{1/2}\zeta),
   \quad \zeta=\exp(\sqrt{2}i\langle M\rangle/f),
  \label{eq:para}
\eeq
where $\langle M\rangle$ represents the condensate, 
$\langle M\rangle=V_+\langle K^+\rangle+V_-\langle K^-\rangle$, with 
$K^{\pm}=(\phi_4\pm i\phi_5)/\sqrt{2}$ and
$\theta^2\equiv 2K^+K^-/f^2$, while $U_f =\exp[2iT_a\phi_a/f]$
means the fluctuation field. It may be interesting to see that this
procedure can be also regarded as the separation of the zero-mode
\cite{gass}.

Accordingly,
defining a new baryon field $B'$ by way of
\beq
  B'=u^\dagger B u,
\eeq
we can see eventually that 
\begin{eqnarray}
  {\cal L}_{chiral}(U, B)
	&=&
  {\cal L}_0(U_f, B')+{\cal L}_{SB}(\zeta U_f\zeta,u B'
  u^\dagger),\nonumber \\
  \delta{\cal L}(U, B)
	&=&\delta{\cal L}(\zeta U_f\zeta,u B'
  u^\dagger).
\label{sbterm}
\end{eqnarray}
Thus only the SB terms {\it prescribe} the $KN$ dynamics in the condensed
phase; we can easily see that
the $KN$ sigma terms, $\Sigma_{Ki} (i=n,p)$, stem from ${\cal
L}_{SB}$, while the Tomozawa-Weinberg term from $\delta{\cal L}$
\cite{TTMY98l}\cite{TTMY}.
The sigma terms can be written by the parameters
in the nonlinear chiral Lagrangian($a_1, a_2, a_3$)
and strange quark mass($m_s$):
\begin{eqnarray}
  \Sigma_{Kp}
	&=& -(a_1+a_2+2a_3)m_s,
	\nonumber \\
  \Sigma_{Kn}
	&=& -(a_2+2a_3)m_s.
\end{eqnarray}
Mass differences of hadrons can determine
$a_1m_s=-67$ MeV and $a_2m_s=134$MeV.
There remains a large uncertainty associated with
the coefficient $a_3m_s$.
Empirically
$a_3m_s$ lies in the range $-134$ to $-310$MeV,
while lattice gauge simulations provide
$a_3m_s = -(220\pm40)$MeV\cite{thorsson}\cite{lattice}.
Later we mainly use $a_3m_s = -222$MeV as a most reasonable value
and add some results for the weak limit $a_3m_s = -134$MeV
to see the parameter dependence.

The integration measure is properly approximated as
$[dU_f]\simeq \prod_{a=1}^8[d\phi_a]$
in the neighborhood of the condensate
and finally we find  
\beq
Z_{chiral}\simeq \int\prod_{a=1}^8[d\phi_a]
[dB'][d\bar B']
\exp\left[\int_0^\beta d\tau\int d^3x
{\cal L}_{chiral}^{\it eff}(\zeta, U_f, B')\right],
\label{effpart}
\eeq
with the effective chiral Lagrangian, 
${\cal L}_{chiral}^{\it eff}(\zeta, U_f, B')$
$=$
${\cal L}_0(U_f, B')$
$+$
${\cal L}_{SB}(\zeta U_f\zeta,u B'u^\dagger)$
$+$
$\delta{\cal L}(\zeta U_f\zeta,u B'u^\dagger)$.

We then evaluate
the partition function $Z_{chiral}$ up to the one-loop
level under the Hartree approximation for kaon-nucleon interactions;
we need not care about the nucleon loops in the heavy baryon limit
(HBL), which we hereafter use in this paper.

During this course, the poles of the 
thermal Green function for kaons 
give the excitation spectra of kaonic modes
with energies $(E_\pm)$.
The mode with $E_-$ is the Goldstone mode and exhibits the Bogoliubov
spectrum,
which stems from the spontaneous breaking of $V$-spin
symmetry in the condensed phase. 
It gives a large contribution
to the thermodynamic quantities
through the Bose-Einstein distribution function.
Under the relevant approximation,
they are reduced to the simple form
\beq
  E_\pm({\bf p})
  \simeq
	\sqrt{p^2+\cos\langle\theta\rangle m_K^{*2}+b^2}
	\pm(b+\cos\langle\theta\rangle\mu_K),
 \label{eq:dispa}
\eeq
where $m_K^{*2} = m_K^2 - \rho_B/f^2 (\Sigma_{Kp}x+\Sigma_{Kn}(1-x))$
and $b=\rho_B(1+x)/(4f^2)$.
We use this form in the following calculations.
We have discussed its relevance in the previous papers\cite{TTMY}
and have seen its practical usefulness
in giving the thermodynamic quantities. 

Eventually the effective thermodynamic potential 
$\Omega_{chiral}=-T\ln Z_{chiral}$ reads
\beq
  \Omega_{chiral}=\Omega_c+\Omega_K^{th}+\Omega_N,
   \label{eq:omega}
\eeq
where $\Omega_c$ is the classical kaon contribution,
\beq
  \Omega_c=V[-f^2m_K^2(\cos\langle\theta\rangle-1)
   -1/2\cdot\mu_K^2f^2\sin^2\langle\theta\rangle],
  \label{eq:omegac}
\eeq
and the thermal kaon contribution, 
is given as follows;
\beq
  \Omega_K^{th}=TV\int \frac{d^3 p}{(2\pi)^3}
  \ln(1-e^{-\beta E_+({\bf p})})(1-e^{-\beta E_-({\bf p})}).
\eeq
It is to be noted that the zero-point-energy contribution of kaons is
very small \cite{TandE}\cite{MYTT}
and we discard it in this paper.


$\Omega_N$ denotes the nucleon contribution.
In order to reproduce the bulk property of nuclear matter
and get a realistic EOS for kaon condensed matter,
we should take into account nucleon-nucleon interactions.
Following Prakash et al.\cite{ains}
we introduce the potential energy in the form,
\beq
  E^{pot}_N = E^{sym}_N + E^{V}_N,
\eeq
where $E^{sym}_N$ represents the symmetry energy contribution
and $E^{V}_N$ the residual potential contribution.
The symmetry energy is given by
\beq
  E^{sym}_N
    = V\rho_B(1-2x)^2 S^{pot}(u); 
\quad
(u = \rho_B/\rho_0; \rho_0 =
0.16{\rm fm}^{-3})
  \label{eq:sym}
\eeq
where the function $S^{pot}(u)$ reads,
\beq
  S^{pot}(u)=(S_0-(2^{2/3}-1)(3/5)E_F^0)F(u)
\eeq
with the constraint $F(u=1)=1$
to reproduce the empirical symmetry energy $S_0\simeq 30$MeV 
at the nuclear density $\rho_0$.
$E_F^0$ is the fermi energy at $\rho_0$.
Hereafter we use $F(u)=u$ for an example. 
It is well-known that the nuclear symmetry energy, which stems from
the kinetic and potential energies, plays an important role for the
ground-state properties of the condensed phase.
Since we have already included 
the kinetic energy for nucleons,
we must take into account only the potential
energy contribution, which should be beyond chiral dynamics. 

On the other hand,
as another potential contribution besides the symmetry energy
we use the following formula
which simulates the results given by sophisticated theoretical
calculations\cite{ains},
\beq
  E^{V}_N/V
    = \frac{1}{2} A u^2 \rho_B 
	+ \frac{B u^{\sigma+1}\rho_B}{1+B'u^{\sigma-1}}
 	+ 3 u \rho_B \sum_{i=1,2} C_i \left(
	\frac{\Lambda_i}{p_F^0} \right)^3
	\left( \frac{p_F}{\Lambda_i}
	- \arctan \frac{p_F}{\Lambda_i} \right).
  \label{eq:vu}	
\eeq
We choose the parameter set
($A$,$B$,$B'$,$\sigma$,$C_i$,$\Lambda_i$)
for compression modulus to be $K_0=240$MeV for normal nuclear matter
\footnote{
Since this parameter set has been also used in Ref.\cite{thorsson}
in which $V(u)$ corresponds to our expression $E^{V}_N$,
our results at zero temperature are the same as theirs.
}. 
Strictly speaking,
the effective potential functions Eqs.(\ref{eq:sym}),(\ref{eq:vu})
should include
temperature dependence by way of the Pauli blocking effect on
nucleon-nucleon scattering in matter.
However, as has been shown in
ref. \cite{TakaHiu88}
temperature-dependence should be weak up to several tens of MeV, 
so that we can use the
form $E_N^{pot}$ even at finite temperature.

Including the potential contribution $E^{pot}_N$,
we get the
single particle energies,
\begin{eqnarray}
  \epsilon_p({\bf p})
    &=& \frac{p^2}{2M} - \left(\Sigma_{Kp}+\mu_K\right)
	\left( 1-\cos\langle\theta\rangle\right)
	+ \frac{1}{V}\frac{\partial E^{pot}_N}{\partial \rho_p},
	\nonumber\\
  \epsilon_n({\bf p})
    &=& \frac{p^2}{2M} - \left(\Sigma_{Kn}+\mu_K/2\right)
	\left(1-\cos\langle\theta\rangle\right)
	+ \frac{1}{V}\frac{\partial E^{pot}_N}{\partial \rho_n}.
  \label{eq:single}
\end{eqnarray}
We have used the non-relativistic approximation for nucleons,
where the scalar density
is not distinguished from the
ordinary density
$\rho_i$.

Then nucleon contributions
can be written in the form;
\beq
  \Omega_N
    = \Omega^{kin}_N + \Omega^{pot}_N, 
\eeq
where $\Omega^{kin}_N$ is
the ``kinetic energy'' contribution
and $\Omega^{pot}_N$ the ``potential energy'' contribution,
\begin{eqnarray}
  \Omega_N^{kin}
    &\simeq& {}-2TV\sum_{n,p}\int\frac{d^3p}{(2\pi)^3}
	\ln(1+e^{-\beta(\epsilon_i({\bf p})-\mu_i)}),
	\nonumber\\
  \Omega^{pot}_N
    &=& {}-\rho_B\frac{\partial E^{pot}_N}{\partial \rho_B}.
	\label{eq:npot}
\end{eqnarray}


Here it is useful to introduce the new parameters $\mu_i^0 (i=n,p)$
instead of the chemical potentials $\mu_i$,
\begin{eqnarray}
 \mu_p^0
  &=& \mu_p + (\mu_K+\Sigma_{Kp})(1-\cos\langle\theta\rangle)
	- \frac{1}{V}\frac{\partial E^{pot}_N}{\partial \rho_p},
	\nonumber\\
 \mu_n^0
  &=& \mu_n + \left(\frac{\mu_K}{2}+\Sigma_{Kn}\right)
	(1-\cos\langle\theta\rangle)
	- \frac{1}{V}\frac{\partial E^{pot}_N}{\partial \rho_n}.
  \label{eq:muo}
\end{eqnarray}
Then $\Omega_N^{kin}$ can be written in the simple form,
\beq
 \Omega_N^{kin}
  \simeq -2TV\sum_{n,p}\int\frac{d^3p}{(2\pi)^3}
	\ln(1+e^{-\beta(\epsilon^0({\bf p})-\mu_i^0)}),
\eeq
with the ``free'' kinetic energy $\epsilon^0({\bf p})=p^2/2M$.

Using the thermodynamic relations,
\beq
 S_{chiral}=-\frac{\partial\Omega_{chiral}}{\partial T}, \quad
  Q_i=-\frac{\partial\Omega_{chiral}}{\partial\mu_i}, \quad
  E_{chiral}=\Omega_{chiral}+TS_{chiral}+\sum\mu_iQ_i,
 \label{eq:rel}
\eeq
we  can find charge, entropy and internal energy. 
The entropy is given by
\begin{eqnarray}
 S_{chiral}
  &=& -\beta\Omega_{chiral}
	+\beta Vf^2\left(-\frac{1}{2}\mu_K^2\sin^2\langle\theta\rangle
	+m_K^2(1-\cos\langle\theta\rangle)\right)
	\nonumber\\
  && {} +\beta V\int\frac{d^3p}{(2\pi)^3}[E_-({\bf p})f_B(E_-({\bf p}))
	+E_+({\bf p})f_B(E_+({\bf p}))]
	\nonumber\\
  && {} +2\beta V\sum_{n,p}\int\frac{d^3p}{(2\pi)^3}
	[(\epsilon^0({\bf p})-\mu_i^0)f_F(\epsilon^0({\bf p})-\mu_i^0)]
	\nonumber\\
  && {} -\beta V(1-2x)^2S^{pot}(u)\rho_B,
 \label{eq:entropy}
\end{eqnarray}
with the Fermi-Dirac distribution function $f_F(E)$ and the
Bose-Einstein distribution function $f_B(E)$.
The kaonic charge is given by,
\beq
 Q_K=V\left[\mu_Kf^2\sin^2\langle\theta\rangle
  +\cos\langle\theta\rangle n_K
  +(1+x)\sin^2(\langle\theta\rangle/2)\rho_B\right],
 \label{eq:qk}
\eeq  
with the number density of thermal kaons $n_K$,
\beq
 n_K=\int\frac{d^3 p}{(2\pi)^3}
  \left[f_B(E_-({\bf p}))-f_B(E_+({\bf p}))\right].
 \label{eq:nk}
\eeq
Nucleon charges are given by
\beq
 Q_i=2V\int\frac{d^3p}{(2\pi)^3}
  \left[f_F(\epsilon^0({\bf p})-\mu_i^0)\right],\quad (i=n,p).
 \label{eq:qn}
\eeq
Then the total energy is given by
\begin{eqnarray}
 E_{chiral}&=&Vf^2\left(-\frac{1}{2}\mu_K^2\sin^2\langle\theta\rangle
  +m_K^2(1-\cos\langle\theta\rangle)\right)
  \nonumber\\
 &&{}+V\int\frac{d^3p}{(2\pi)^3}[E_-({\bf p})f_B(E_-({\bf p}))
  +E_+({\bf p})f_B(E_+({\bf p}))]\nonumber\\
 &&{}+2V\sum_{n,p}\int\frac{d^3p}{(2\pi)^3}
  \left[\left( \epsilon_i({\bf p})
  - \frac{1}{V}\frac{\partial E^{pot}_N}{\partial \rho_n} \right)
  f_F(\epsilon^0({\bf p})-\mu_i^0)\right]
  \nonumber\\
 &&{}+V(1-2x)^2S^{pot}(u)\rho_B+\mu_KQ_K,
  \label{eq:energy}
\end{eqnarray}
where we add the term with
$\displaystyle
\frac{1}{V}\frac{\partial E^{pot}_N}{\partial \rho_n}$,
to avoid double-counting.

The total thermodynamic potential  $\Omega_{total}$ 
is given by adding the one for
leptons (electrons, muons and neutrinos),
$\Omega_l$, $\Omega_{total}=\Omega_{chiral}+\Omega_l$; 
\beq
\Omega_l=-2TV\sum_{e,\mu, \nu}\int\frac{d^3
p}{(2\pi)^3}\left[\ln(1+e^{-\beta(\varepsilon_i({\bf p})-\mu_i)})
+\ln(1+e^{-\beta(\varepsilon_{\bar i}({\bf p})+\mu_i)})\right],
\label{omegal}
\eeq
with $\varepsilon_i({\bf p})=\varepsilon_{\bar i}({\bf
p})=\sqrt{m_i^2+p^2}$
($i=e,\mu,\nu_e,\nu_\mu$).

The parameters $x,\langle\theta\rangle$ and chemical potentials 
$\mu_K,\mu_i(i=n,p)$
are determined by the extremum conditions for given density and
temperature.
The first condition demands $\partial\Omega_{total}/\partial
x=0$, which is equivalent with the condition for chemical equilibrium
among kaons, nucleons:
\beq
 \mu_n-\mu_p=\mu_K,
 \label{eq:chem}
\eeq
or
\beq
 \mu_p^0-\mu_n^0-4S^{pot}(u)(1-2x)+\frac{1-\cos\langle\theta\rangle}{2}
  \{2(\Sigma_{Kn}-\Sigma_{Kp})-\mu_K\}+\mu_K=0,
 \label{eq:che}
\eeq
by way of Eq.(\ref{eq:muo}).
The charge neutrality demands $\partial\Omega_{total}/\partial
\mu_K=0$:
\beq
 Q_K+Q_e+Q_\mu=Q_p(=xV\rho_B),
 \label{eq:cha}
\eeq
where the hadron charges are given in Eqs.~(\ref{eq:qk}),(\ref{eq:qn}) and 
the lepton numbers are given by 
\begin{eqnarray}
Q_i\equiv Y_i\rho_B=2V\int\frac{d^3
p}{(2\pi)^3}\left[f_F(\epsilon_i({\bf p})-\mu_i)
-f_F(\epsilon_{\bar i}({\bf p})+\mu_i)\right],
\end{eqnarray}
with the lepton-number fractions $Y_i$ ($i=e,\mu$).
Finally the extremum condition
with respect to $\langle\theta\rangle$ gives
\beq
 0 = f^2\sin\langle\theta\rangle
	(m_K^{*2}-2\mu_Kb-\mu_K^2\cos\langle\theta\rangle)
	+\frac{\partial(\Omega_K/V)}
	{\partial\langle\theta\rangle},
 \label{eq:feq}
\eeq
which is just the field equation of motion for kaon at momentum zero. 
Eqs.(\ref{eq:che}),(\ref{eq:cha}), (\ref{eq:feq}) 
are the basic equations to determine the equilibrated state.

We can see that once the fluctuation 
contribution (the second term in Eq.(\ref{eq:feq})) is taken into
account, 
the relation 
$E_-({\bf p}=0)=0$ no longer holds. So the Goldstone nature is
violated up to one-loop order in the loop expansion.
This situation always occurs 
in the perturbation theory \cite{kap}: 
the number of loops are {\it always}
different between the thermodynamic potential ($N_l$) and the self-energy
for kaons ($N_l-1$). This shortcoming is never cured
unless we perform a full order calculation. 
This matter is beyond the scope of this paper.
Fortunately, 
the contribution of  fluctuations
can be estimated to be small in the
temperature-density region we are interested in\cite{TTMY}.

\section{Numerical Results and Discussions}\label{sect:nume}

Using the formulation given in Sect.\ref{formulation},
we study the properties of kaon condensed matter
taking into account thermal and neutrino-trapping effects
and then discuss the properties of a PNS.
Finally
some implications on its delayed collapse
will be given.
In this paper, we use $a_3m_s=-222$MeV
($\Sigma_{KN}=\frac{1}{2}(\Sigma_{Kp}+\Sigma_{Kn})=344$ MeV)
which is known as a most reasonable value
(as discussed in Sect.\ref{formulation}).
In Sect.\ref{sec:a3ms134}
we will give some results in the case of weak coupling limit,
$a_3m_s=-134$MeV corresponding to $\Sigma_{KN}=168$ MeV
to study the dependence of the EOS
on the strength of the $KN$ interaction.

\subsection{Phase Diagram}
\begin{figure}[htb]
  \epsfsize=0.6\textwidth
  \epsffile{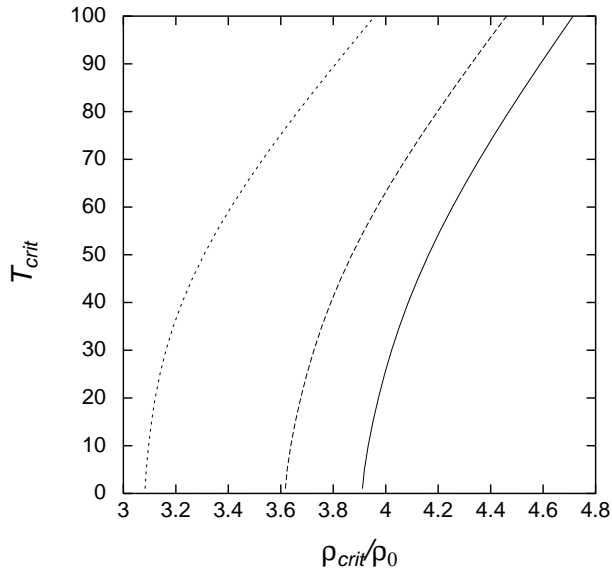}
 \caption{Phase Diagram:
solid line and dashed line
show the critical lines
in the neutrino-trapped cases, $Y_{le}=0.4$ and $0.3$,
respectively
and dotted line in the neutrino-free case.
}
 \label{fig:pd}
\end{figure}

First we show the phase diagram  in Fig.\ref{fig:pd}
for neutrino-free and trapping matter.
In the neutrino-trapping case,
there exist additional conditions
with respect to the conservation of lepton numbers\cite{thorsson}.
Recent calculations of gravitational core-collapse
of massive stars
indicate the electron lepton number $Y_{le}$
at the onset of neutrino-trapping $Y_{le}\approx 0.35$\cite{Epst81}.
Then we take
$Y_{le}\equiv Y_e+Y_{\nu_e}=0.4$ and $0.3$
and for muon
$Y_{l\mu}\equiv Y_\mu+Y_{\nu_\mu}=0.0$
in the neutrino-trapping case.
On the other hand, in the neutrino-free case,
$Y_{\nu_e}=Y_{\nu_\mu}=0$.

In the phase diagram the temperature-density 
plane is separated into two regions by the
critical line;
the left-side region means normal phase,
and the right-side region the kaon-condensed phase.
We can see that both of the neutrino-trapping and thermal effects
work against the phase transition. 
In the low-temperature case the neutrino-trapping
effect is dominant, while both effects become comparable at
high-temperature. 

The neutrino-trapping effect
may be simply understood as follows.
Kaon condensation occurs when the lowest kaon energy ($\epsilon_{K^-}$) 
is equal to the kaon chemical potential ($\mu_{K}$):
$\epsilon_{K^-}$ decreases by attractive 
kaon-nucleon interactions as density increases, 
while $\mu_{K}=\mu_n - \mu_p $ increases
to match $\epsilon_{K^-}$ at the critical density.
In the neutrino-trapping case,
the chemical equilibrium due to weak interactions
implies that
the relation,
$\mu_{K} = \mu_n - \mu_p = \mu_e - \mu_{\nu_e}$, holds.
Compared with the neutrino-free case where
$\mu_{\nu_e}=0$,
$\mu_{K}$ should be reduced
due to the finite contribution of $\mu_{\nu_e}$.
Hence higher density is needed to satisfy
the condition $\epsilon_{K^-}=\mu_K$.


\subsection{EOS for thermal and isentropic matter}\label{sec:eos}

\begin{figure}[htb]
 \vspace{2mm}
 \begin{minipage}{0.49\textwidth}
  \epsfsize=0.99\textwidth
  \epsffile{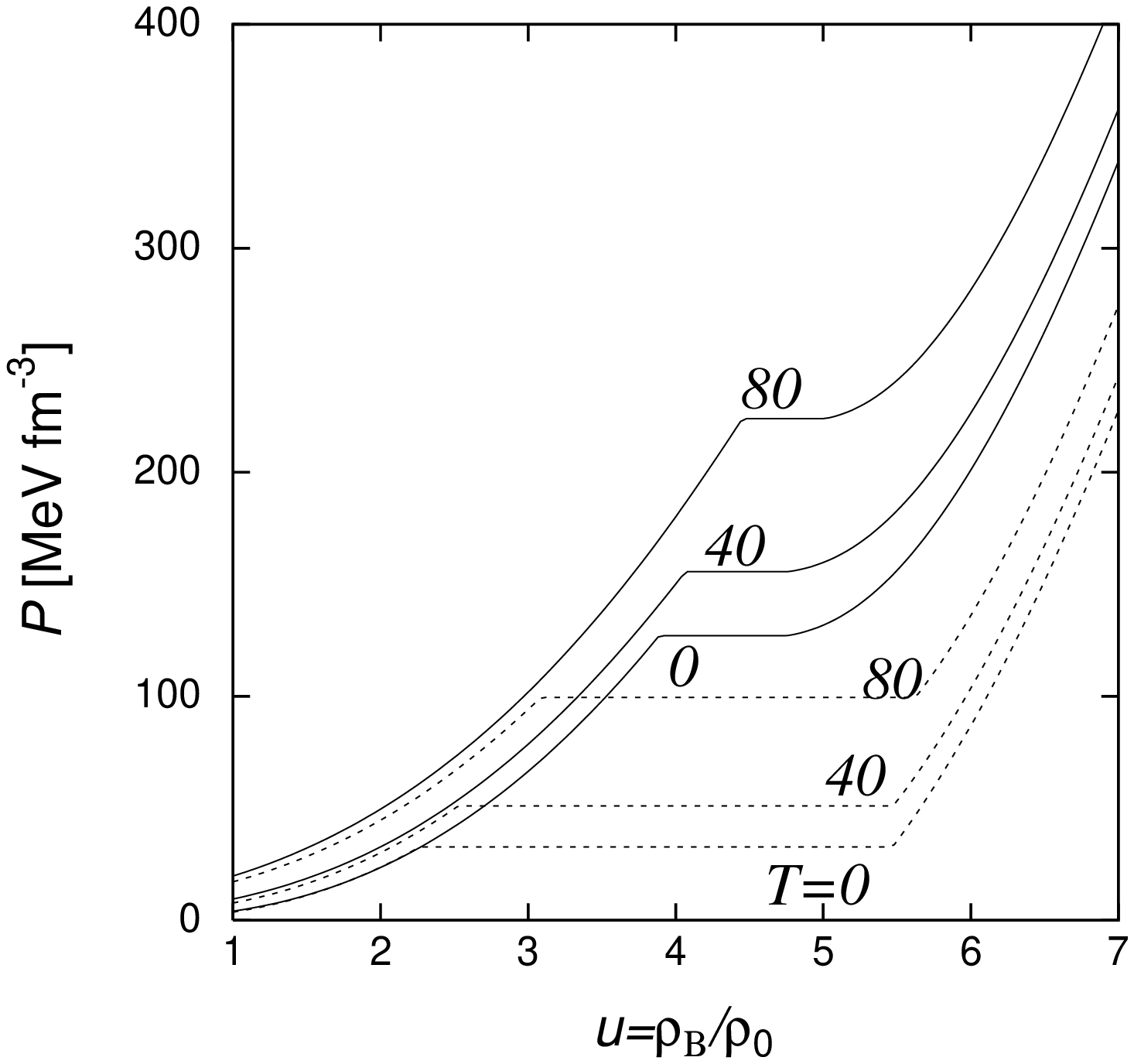}
 \end{minipage}%
 \hfill~%
 \begin{minipage}{0.49\textwidth}
  \epsfsize=0.99\textwidth
  \epsffile{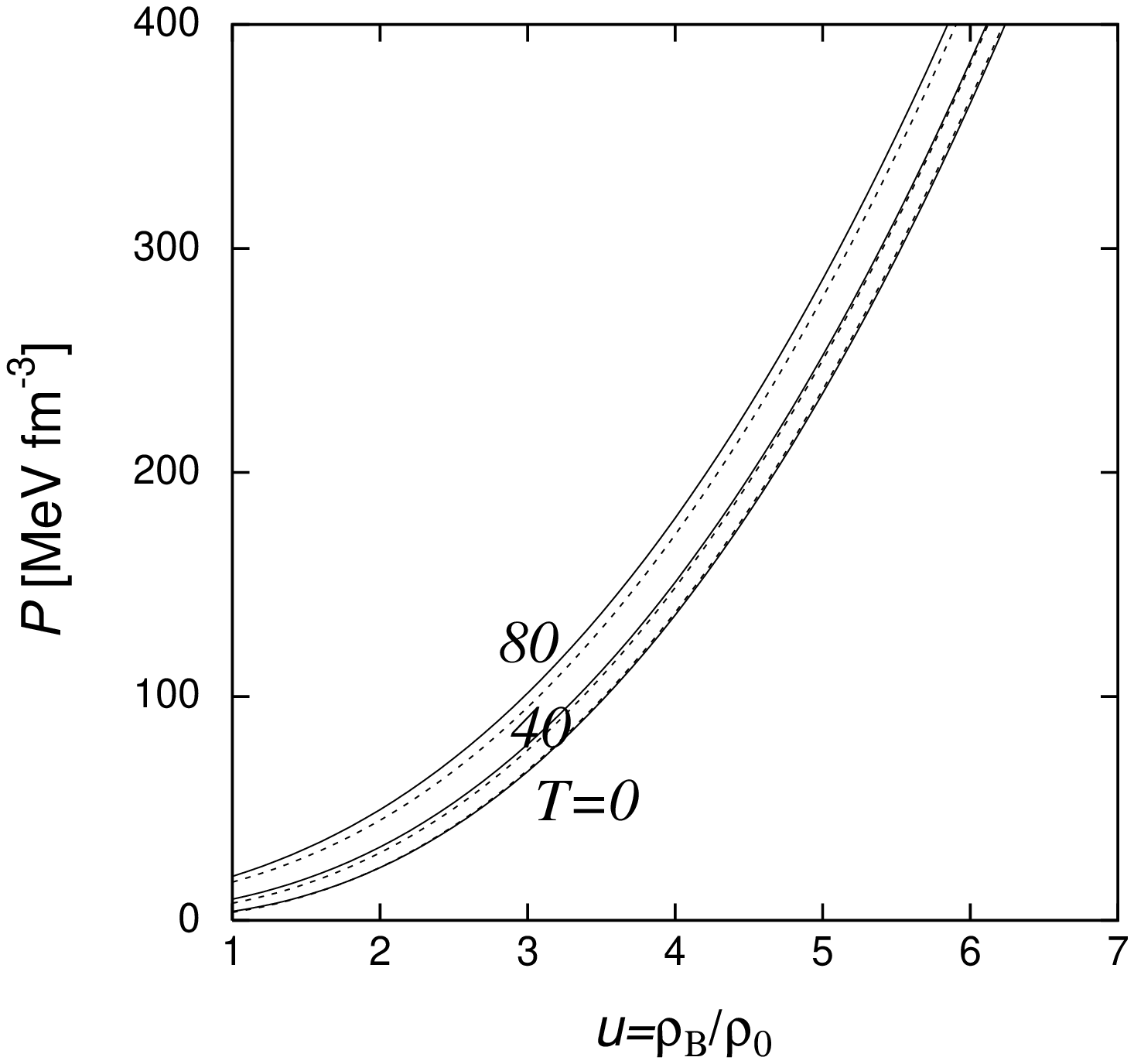}
 \end{minipage}%
 \caption{Pressure of isothermal matter ($T=0,40,80$MeV)
for 
kaon condensed matter[left panel]
and 
for normal nuclear matter in beta-equilibrium [right panel].
Solid lines and dashed lines are in the neutrino-trapped case
($Y_{le}=0.4$) and
the neutrino-free case, respectively.}
 \label{fig:isoTuPP}
\end{figure}

Here
we compare the equations of state in the two cases: the isothermal and
isentropic cases.
The equations of state
under the isothermal condition ($T=0, 40$ and $80$ MeV)
for kaon condensed matter and normal matter
are depicted
in Fig.\ref{fig:isoTuPP}.
The resulting EOS for kaon condensed matter
shows the existence of
the thermodynamically unstable region
because the phase transition is of first order
and large softening of EOS
so that 
we have recourse to the Maxwell construction to obtain
the realistic EOS in equilibrium\footnote{
Strictly speaking,
we need to apply the Gibbs conditions
because there exist two chemical potentials
\cite{gleGENE}\cite{gle}.
However, we immediately see that the Gibbs conditions cannot be satisfied
in the chiral model(Appendix A).
So we use here the Maxwell construction for simplicity
to get the EOS in equilibrium
and leave this matter for a future study.
}.
There appears the equal-pressure region 
as a result of the Maxwell construction.
We can see that both the thermal
and neutrino-trapping effects stiffen the EOS
mainly through suppression of kaon condensate.
Then,
in comparison of the EOS with that for normal matter,
the thermal effect is more profound,
in particular around the critical density.

The equal-pressure region 
becomes narrower as temperature increases
or more neutrinos are trapped,
which implies that the
magnitude of the first-order phase transition is reduced by
both effects of temperature and neutrinos.
We can see the remarkable narrowing,
especially in the neutrino-trapping case
because trapped neutrinos suppress
not only the occurrence of kaon condensation
but also the growth of the condensate.
We shall see later that the existence of
the equal-pressure region may lead to the gravitationally unstable region
in the branch of neutron stars.

\begin{figure}[htb]
  \epsfsize=0.6\textwidth
  \epsffile{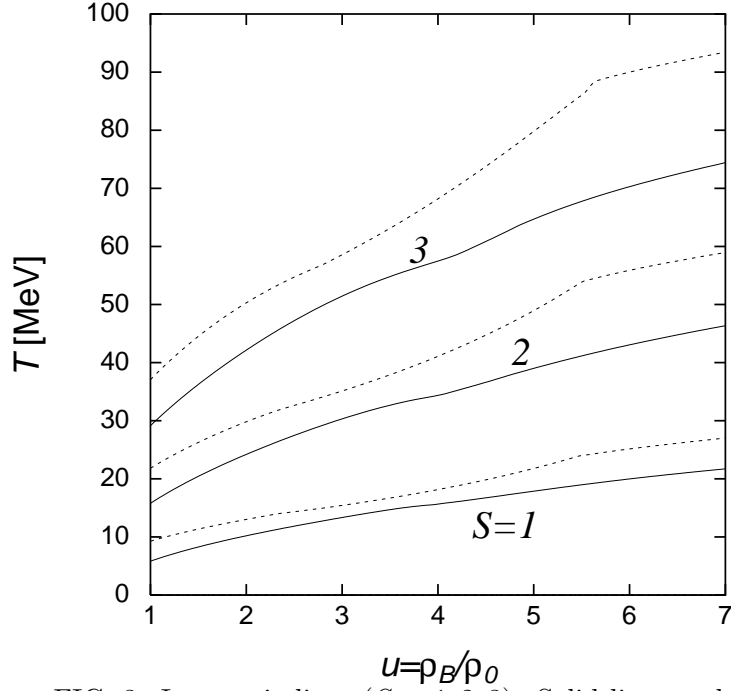}
 \caption{Isentropic lines ($S=1,2,3$).
Solid lines and dashed lines are in the neutrino-trapped case
($Y_{le}=0.4$),
the neutrino-free case, respectively.}
 \label{fig:isoSuT}
\end{figure}

The isentropic EOS might be more relevant for the PNS matter.
First of all we show the isentropic lines in density-temperature plane
in Fig.\ref{fig:isoSuT}
by calculating entropy by the use of Eq.(\ref{eq:entropy})
and lepton contribution for given temperature and density.
In each isentropic line,
nucleons dominantly contribute
to entropy\cite{Prakash}\cite{PonsH}\cite{Ponsprep}.
Under the isentropic condition,
in which entropy per baryon ($S$) is taken
to be constant ($1$,$2$ or $3$)
over the star interior,
temperature increases as density becomes high.
This means that
temperature becomes the highest at the center
of the star
once we construct a PNS
with isentropic matter.

\begin{figure}[htb]
  \epsfsize=0.6\textwidth
  \epsffile{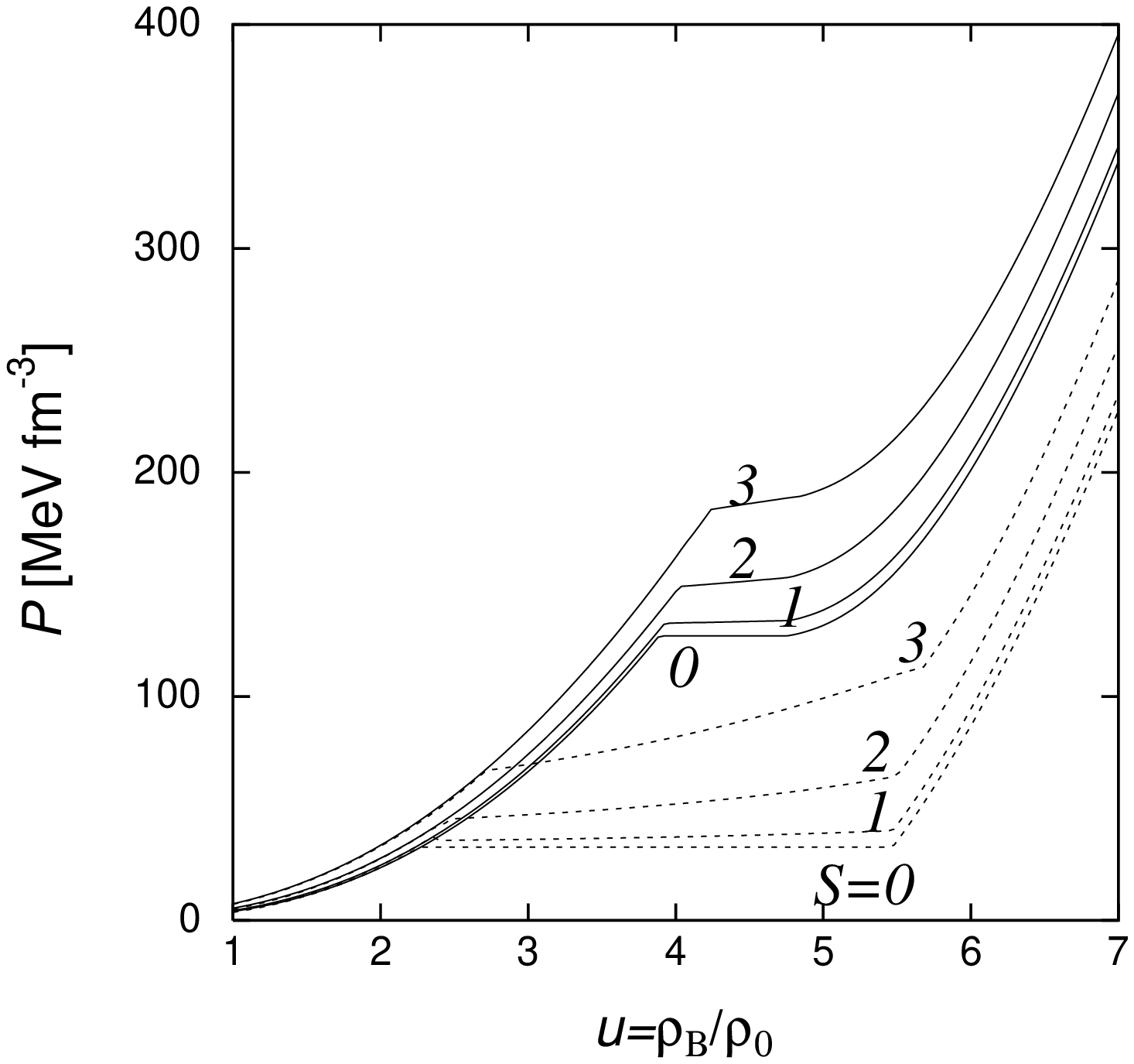}
  \caption{Pressure for isentropic matter ($S=0,1,2,3$).
Solid line and dashed line are in neutrino-trapping case
($Y_{le}=0.4$) and
neutrino-free case respectively.}
 \label{fig:isoSuPP}
\end{figure}

Then the isentropic EOS can be obtained by connecting the values of
pressure at the corresponding
temperatures for given densities (Fig.\ref{fig:isoSuPP}).
It is to be noted that the equal-pressure region, 
under the isentropic condition,
disappears except the $S=0$ cases.

It is well-known that when the equal-pressure region exists in the EOS,
neutron star also has a density gap\cite{Fujii96}.
In the isentropic case,
the equal-pressure region no longer exists
and
the density gap disappears
even if we apply the Maxwell construction for the isothermal EOS.
Instead the mixed phase appears inside the neutron star
\footnote{Note that this mixed phase should be
distinguished from the one resulted from the Gibbs conditions.}.

\subsection{Properties of a PNS
and the possibility of its delayed collapse}
\begin{figure}[htb]
 \vspace{2mm}
 \begin{minipage}{0.49\textwidth}
  \epsfsize=0.99\textwidth
  \epsffile{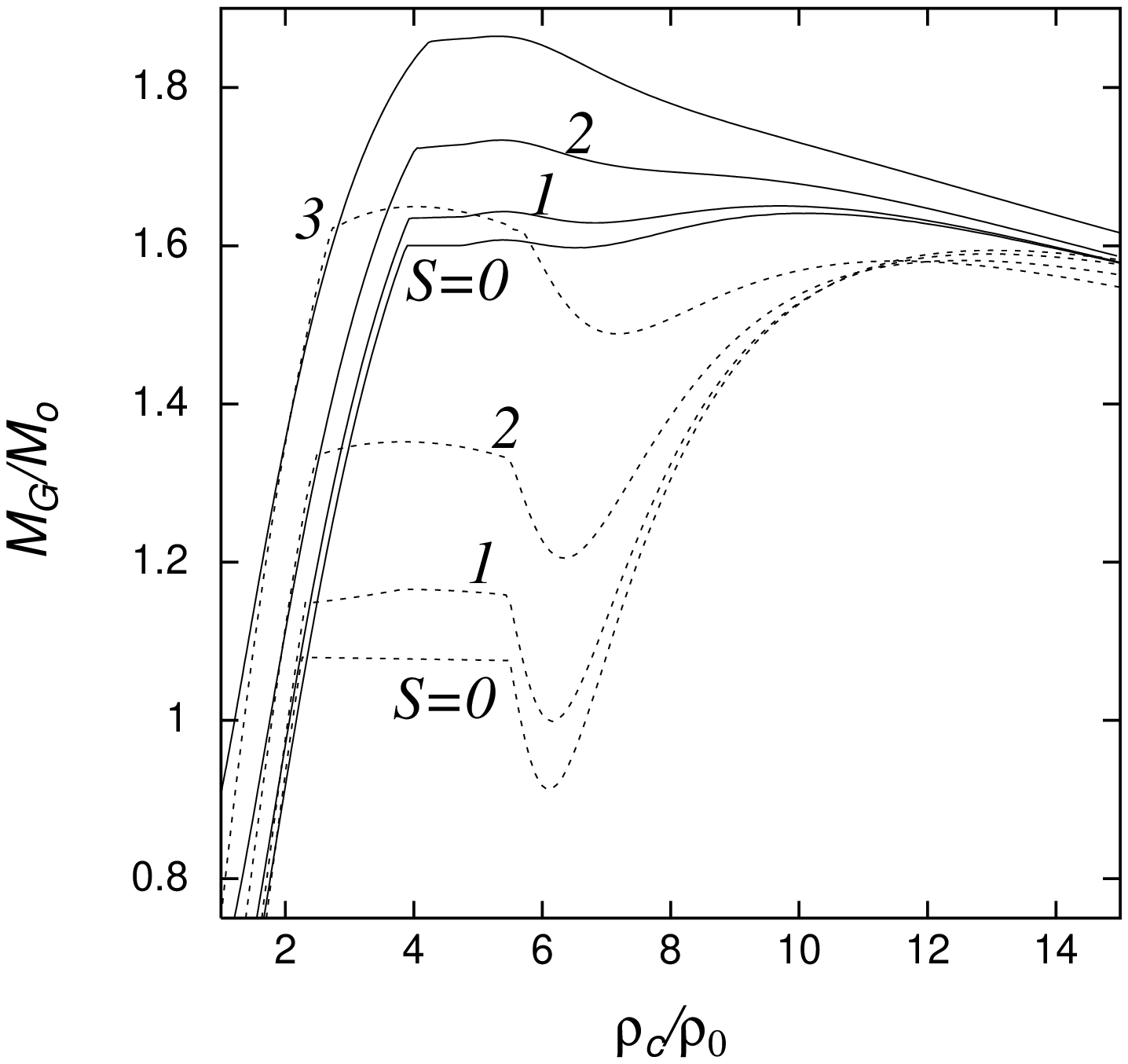}
 \end{minipage}%
 \hfill~%
 \begin{minipage}{0.49\textwidth}
  \epsfsize=0.99\textwidth
  \epsffile{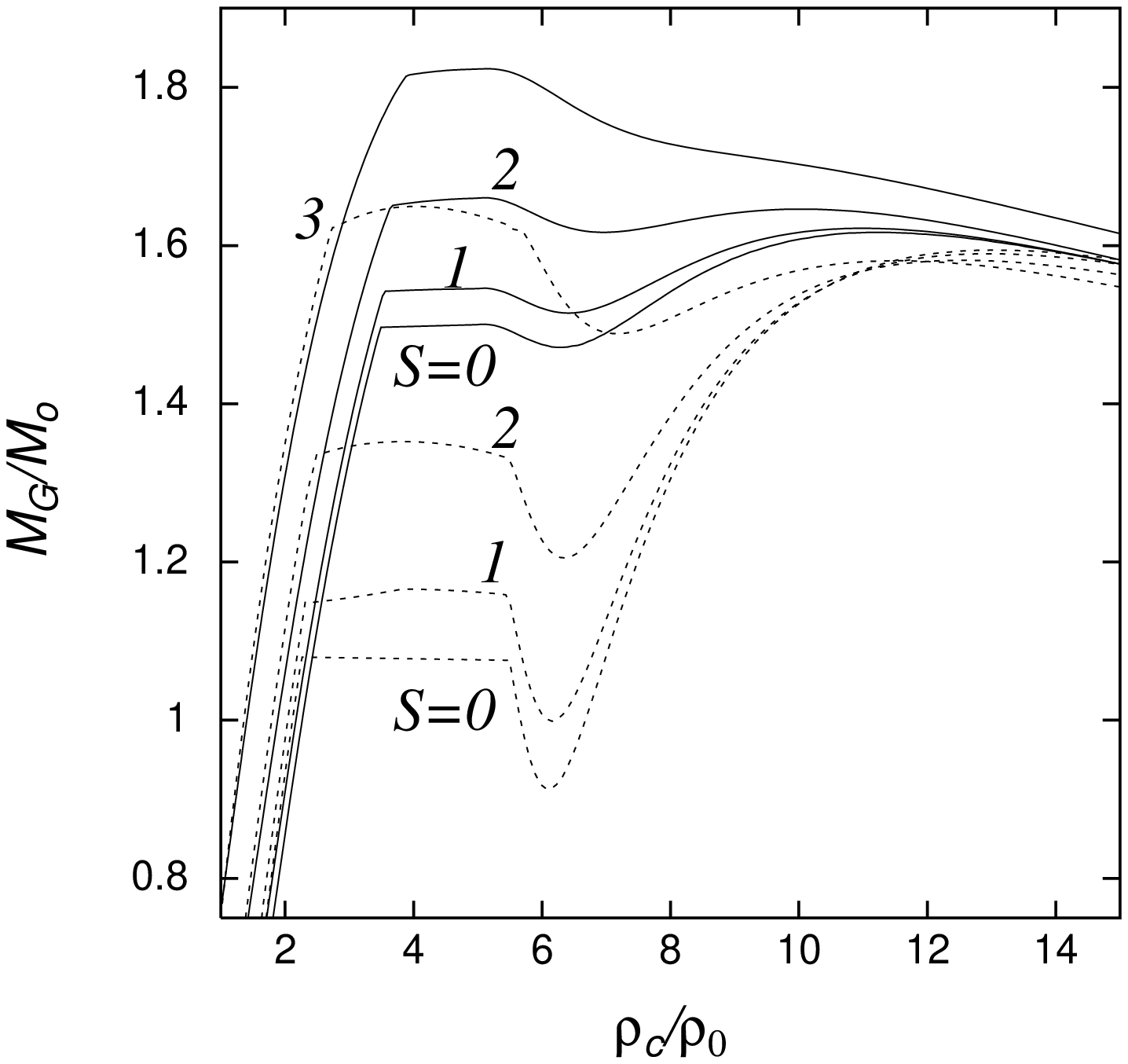}
 \end{minipage}%
 \caption{Central density versus gravitational mass.
Solid lines and dashed lines are in
neutrino-trapping ($Y_{le}=0.4$[left panel] or $Y_{le}=0.3$[right panel]) 
and neutrino-free cases
respectively at $S=0,1,2,3$.
}
 \label{fig:MuP}
\end{figure}

Using the isentropic EOS,
we construct a PNS
by solving TOV equation\footnote{
In the low density region $\rho_B < \rho_0$,
we use the EOS calculated by Lattimer et al\cite{lowEOS}.
}.
In Fig.\ref{fig:MuP}
we show the central density-gravitational mass relation
in the neutrino-free or trapping case with
entropy per baryon $S=0, 1, 2$ or $3$.
We can see
the thermal effect is large
for the stars
whose central density is around the critical density
because the pressure rapidly increases
in this region as temperature does.
On the other hand,
mass of the stars with high central density
seems to be hardly changed by the thermal effect.
This is because
the thermal effect causes not only higher pressure
but also larger energy
which in turn results in the stronger gravity.
Sometimes this leads to
the lower maximum-mass for the higher entropy.

We can find that,
clearly in the neutrino-free case,
once kaon condensation occurs in the core of a star,
gravitationally unstable region (negative gradient part) appears
and the neutron star branch is separated into two stable branches:
one is for stars with kaon condensate in their cores
and the other consisting of only normal matter.
The maximum mass
of the stars which include kaon condensate in the core,
is hardly changed 
but decreases by the thermal effect
because of the strong gravity
except the $S=3$ case.

In the $S=0,1$ and neutrino-trapping cases,
we can see that the neutron-star branch is also separated by the
gravitationally unstable region and the maximum-mass 
lies on
the branch with kaon condensate.

On the other hand,
in other cases: the $S=3$ and neutrino-free case,
the $S=2,3$ and neutrino-trapping cases,
almost all of the stars 
with kaon condensate are gravitationally unstable,
and the maximum-mass star still
resides on the normal branch.
For this reason
the central density of the maximum-mass star 
is very different from those in the $S=0$ or $1$ case.
In table \ref{table:mm}
we summarize the data for stars with maximum-mass
in each case.
\begin{table}[ht]
\begin{tabular}{rcccc}
 		&branch	& central density & gravitational mass & total baryon number\\ \hline
	\multicolumn{5}{l}{\underline{$\nu$-free $Y_{\nu_e}=0$}} \\
	~ ~ ~ ~ $S=0$ & K 	& 13.2$\rho_0$ & $1.59M_\odot$ & $2.24\times 10^{57}$ \\
	~ $S=1$	& K	& 13.5$\rho_0$ & $1.59M_\odot$ & $2.22\times 10^{57}$ \\
	~ $S=2$	& K	& 12.5$\rho_0$ & $1.58M_\odot$ & $2.14\times 10^{57}$ \\
	~ $S=3$	& N	& 4.0$\rho_0$  & $1.65M_\odot$ & $2.11\times 10^{57}$ \\ \hline
	\multicolumn{5}{l}{\underline{$\nu$-trapping $Y_{le}=0.4$}} \\
	~ $S=0$ & K 	& 10.0$\rho_0$ & $1.64M_\odot$ & $2.13\times 10^{57}$ \\
	~ $S=1$	& K	& 9.6$\rho_0$  & $1.65M_\odot$ & $2.14\times 10^{57}$ \\
	~ $S=2$	& N	& 5.4$\rho_0$  & $1.73M_\odot$ & $2.23\times 10^{57}$ \\
	~ $S=3$	& N	& 5.3$\rho_0$  & $1.86M_\odot$ & $2.36\times 10^{57}$ \\ \hline
	\multicolumn{5}{l}{\underline{$\nu$-trapping $Y_{le}=0.3$}} \\
	~ $S=0$ & K 	& 11.3$\rho_0$ & $1.62M_\odot$ & $2.15\times 10^{57}$ \\
	~ $S=1$	& K	& 11.0$\rho_0$ & $1.62M_\odot$ & $2.14\times 10^{57}$ \\
	~ $S=2$	& N	& 5.1$\rho_0$  & $1.66M_\odot$ & $2.15\times 10^{57}$ \\
	~ $S=3$	& N	& 5.1$\rho_0$  & $1.82M_\odot$ & $2.33\times 10^{57}$ \\
\end{tabular}
 \caption{Configuration for stars with maximum-mass:
which branch the star resides in,
normal branch(N) or kaon condensed branch(K),
its central density, mass and total baryon number.
}
 \label{table:mm}
\end{table}%


To discuss the possibility of
the delayed collapse of the PNS,
the total baryon number $N_B$ should be fixed
as a conserved quantity during the evolution\cite{takatsuka},
under the assumption that there is no accretion
in the deleptonization and initial cooling eras.

\begin{figure}[htb]
 \vspace{2mm}
 \begin{minipage}{0.49\textwidth}
  \epsfsize=0.99\textwidth
  \epsffile{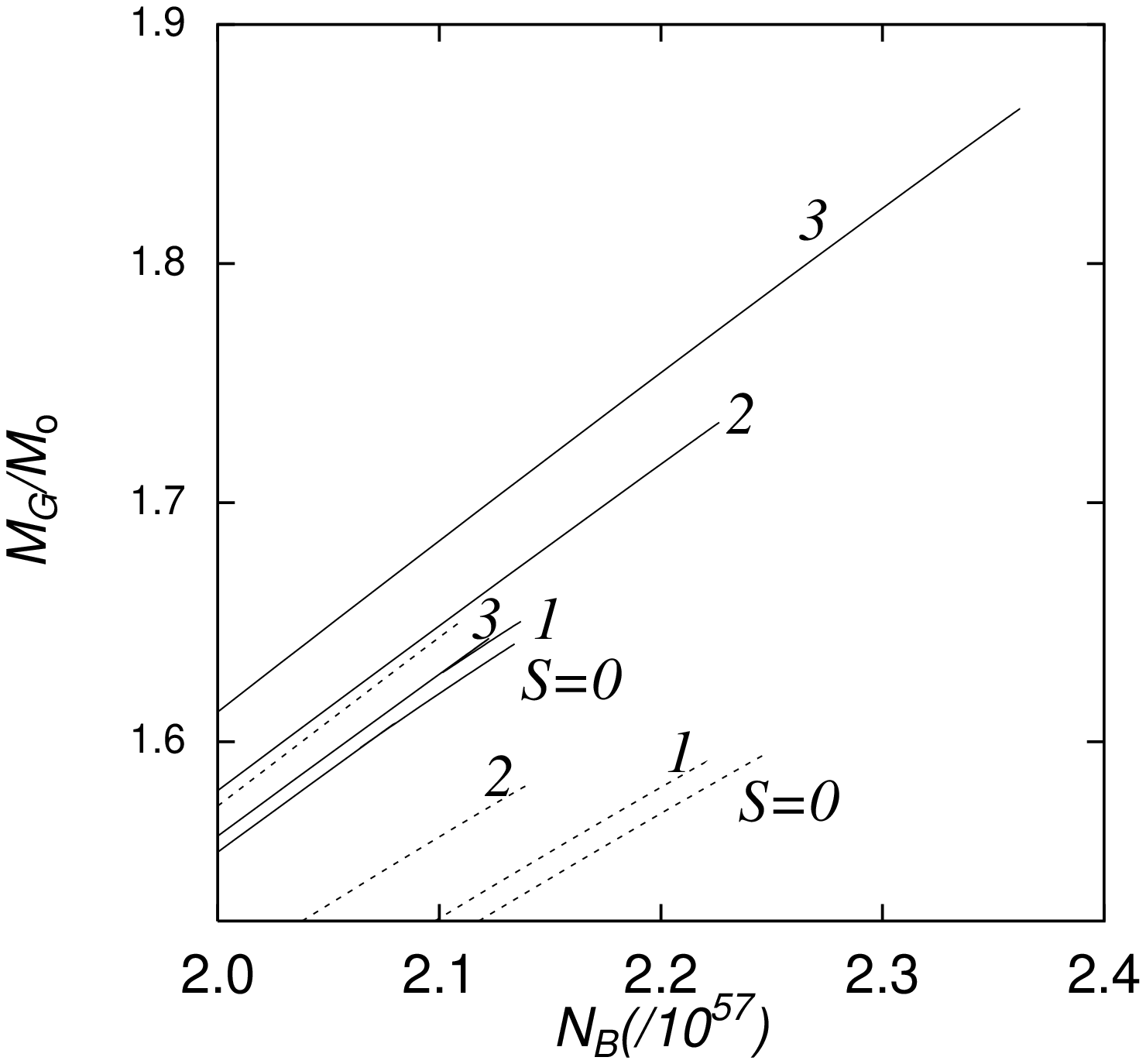}
 \end{minipage}%
 \hfill~%
 \begin{minipage}{0.49\textwidth}
  \epsfsize=0.99\textwidth
  \epsffile{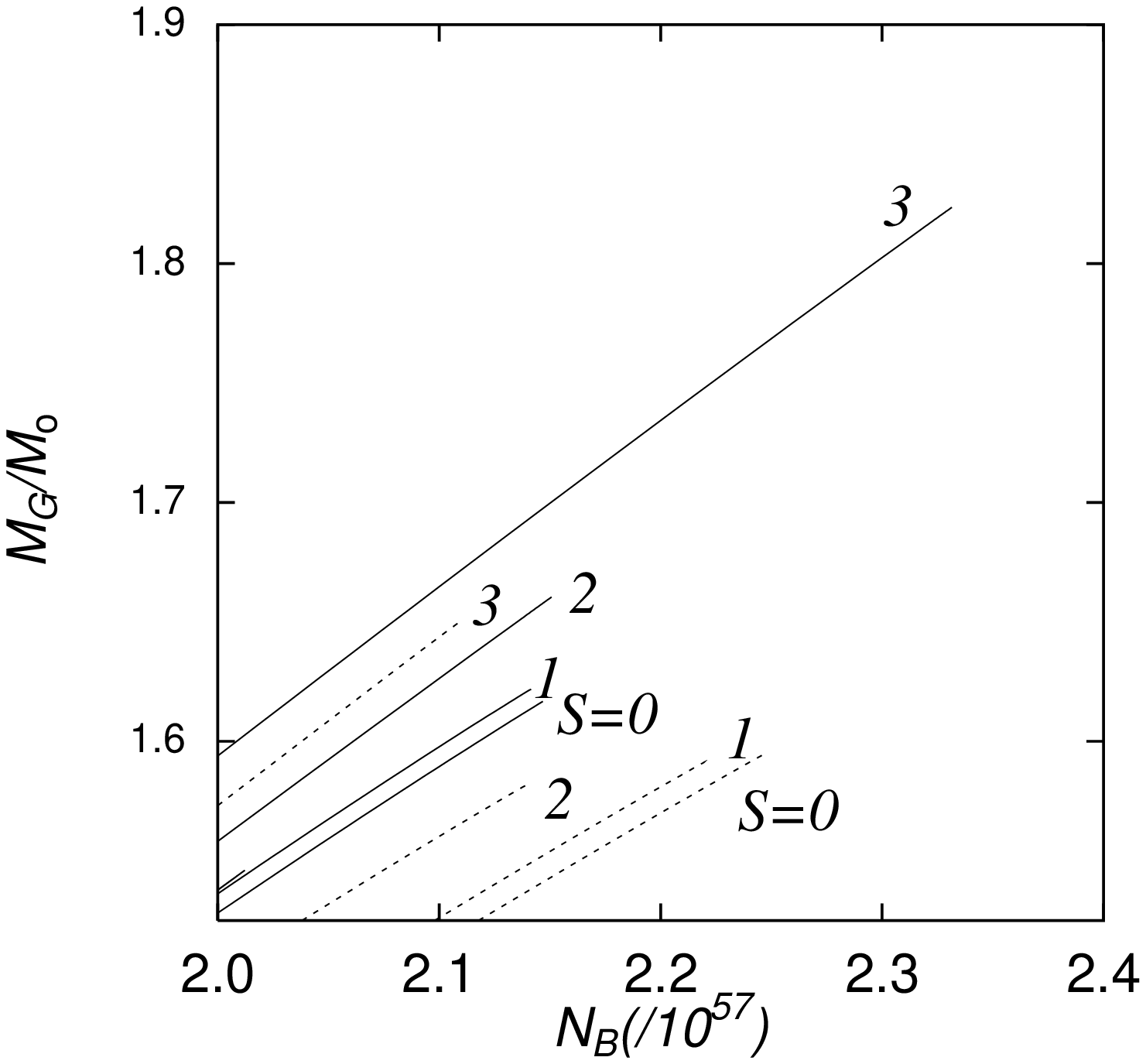}
 \end{minipage}%
 \caption{Total baryon number and gravitational mass
for stable PNS.
Solid lines and dashed lines
in neutrino-trapping ($Y_{le}=0.4$[left panel]
or $Y_{le}=0.3$[right panel]) and
neutrino-free cases,
respectively at $S=0,1,2,3$.
}
 \label{fig:NBMP}
\end{figure}

Discarding gravitationally unstable stars,
we show mass of stable stars
for given baryon numbers
in Fig.\ref{fig:NBMP}.
Each terminal point represents 
the star with the maximum-mass and the maximum total baryon number
in each case.
If the initial mass is beyond the terminal point
when the neutron star is born,
it should collapse into a black hole
(not a delayed collapse but the
usual formation of a black hole during supernova explosion).
We have shown the neutrino-trapping
and free cases;
the former corresponds to the initial stars
before the deleptonization era,
while the latter
to the initial cooling era
after deleptonization.
It is interesting to see the
difference between
the neutrino-free and trapping cases:
the curve is shortened as entropy increases
in the former case, 
while elongated in the latter case.
The difference results from which branch
the maximum-mass star resides in.
These features are
responsible for the following discussion
about the delayed collapse and
the maximum-mass of cold neutron stars. 

The delayed collapse takes place
if the initially stable star
on a curve finds no corresponding stable point
on other curves at any stage
during its evolution through deleptonization or cooling.
When we discuss the evolution,
several time-scales are important:
$t_K$ for the onset and growth of kaon condensation,
$t_{del}$ and $t_{cool}$
for deleptonization and initial cooling eras, respectively.
We can neglect $t_K$
because $t_K \ll$ several sec. $\approx t_{del}$
$<$ 20-30 sec. $\approx$
$t_{cool}$ \cite{Prakash}\cite{MutoTI}.

Now, consider the typical evolution:
for example, a PNS
has initially $Y_{le}=0.4$ and $S=2$ 
after supernova explosion
and evolves
through deleptonization 
into the $S=2$ and neutrino-free stage
as an intermediate stage.
We can see clearly the PNS with 
$1.68M_\odot \le M_G \le 1.73M_\odot$
can exist as meta-stable star at the beginning but
cannot find any point on the curve in the $S=2$ and neutrino-free case.
Therefore it must collapse to the low-mass black hole
in the deleptonization era.
It is to be noted that because the stars
on the curve in the $S=2$ and $Y_{le}=0.4$ case
hardly include kaon condensate,
their collapse are largely due to the appearance of 
kaon condensate in their cores.

Furthermore, we can also determine maximum-mass of 
cold neutron stars
by taking into account the evolution of the PNS,
as pointed out by Takatsuka\cite{takatsuka}.
Usually we assign maximum-mass of cold
neutron stars as the terminal point
on the curve in the $S=0$ and neutrino-free case
in Fig.{\ref{fig:NBMP}
($M_{max}=1.59M_\odot$).
However, it is
wrong when we take into account the evolution of neutron stars,
especially in the initial cooling era. 
In order to see how to determine the realistic
maximum-mass of cold neutron stars,
we consider a typical evolution
with the total baryon number fixed.
During the evolution we have already adopted,
stars with more total baryon number
than the terminal point of the curve
in the $S=2$ and neutrino-free case
should collapse to black-holes
in the deleptonization era
and cannot evolve stably
to the usual cold neutron stars.
Therefore
only stars with the total baryon number
$N_B \le 2.14\times 10^{57}$
can evolve to cold neutron stars
and the corresponding maximum-mass is $1.54M_\odot$
(See Fig.\ref{fig:NBMP} and Table \ref{table:mm}.).
Generally speaking,
the maximum-mass should be determined by
taking into account the evolution\cite{takatsuka}.
In our calculation
it seems that
the neutrino-free and hot stage
plays an important role
to determine the maximum-mass.

If we consider other scenarios,
we can of course
determine the maximum-mass of cold neutron stars
and stars which should collapse to the low mass blacks hole in each case.

We can conclude that
the delayed collapse possibly takes place
in the deleptonization era
due to the occurrence of kaon condensation
and the maximum-mass should be determined
in the initial cooling era.

Next,
we show that
the delayed collapse in the case of the PNS consisting of
only normal matter
(without any phase transition)
is impossible.
In Fig.\ref{fig:NBMnm}
the gravitational mass for gravitationally stable
{\it normal} neutron stars
is plotted as a function of total baryon number.
Each terminal point represents
the maximum-mass star in each configuration.
We can similarly study the stability of neutron stars
in the deleptonization or cooling stage.
After all every star can find
the stable point in each stage,
and the delayed collapse of {\it normal}
neutron stars is impossible.
\begin{figure}[htb]
 \epsfsize=0.49\textwidth
 \epsffile{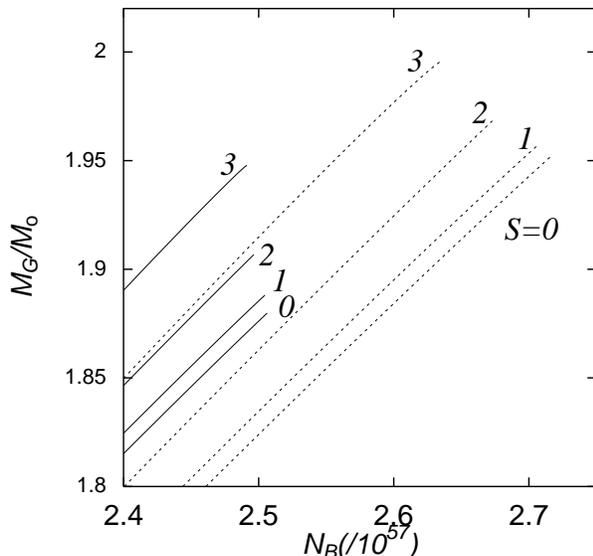}
 \caption{Total baryon number and gravitational mass
for normal matter.
Solid lines and dashed lines are
in neutrino-trapping ($Y_{le}=0.4$)
and neutrino-free cases,
respectively
at $S=0,1,2,3.$
}
 \label{fig:NBMnm}
\end{figure}

\subsection{Case of the weak $KN$ sigma term}\label{sec:a3ms134}

The left panel in Fig.\ref{fig:134} shows the isentropic EOS
for neutrino-free and trapping ($Y_{le}=0.4$) matter
in the case of $\Sigma_{KN}=168$MeV.
The phase transition is of the second order in this case.
The critical density is moved higher
compared to that in the case of $\Sigma_{KN}=344$MeV
(See Fig.\ref{fig:isoSuPP}),
which leads to the weaker softening of the EOS.
There is no thermodynamically unstable region in the isothermal EOS
and we don't need the Maxwell nor the Gibbs constructions.
\begin{figure}[htb]
 \vspace{2mm}
 \begin{minipage}{0.49\textwidth}
  \epsfsize=0.99\textwidth
  \epsffile{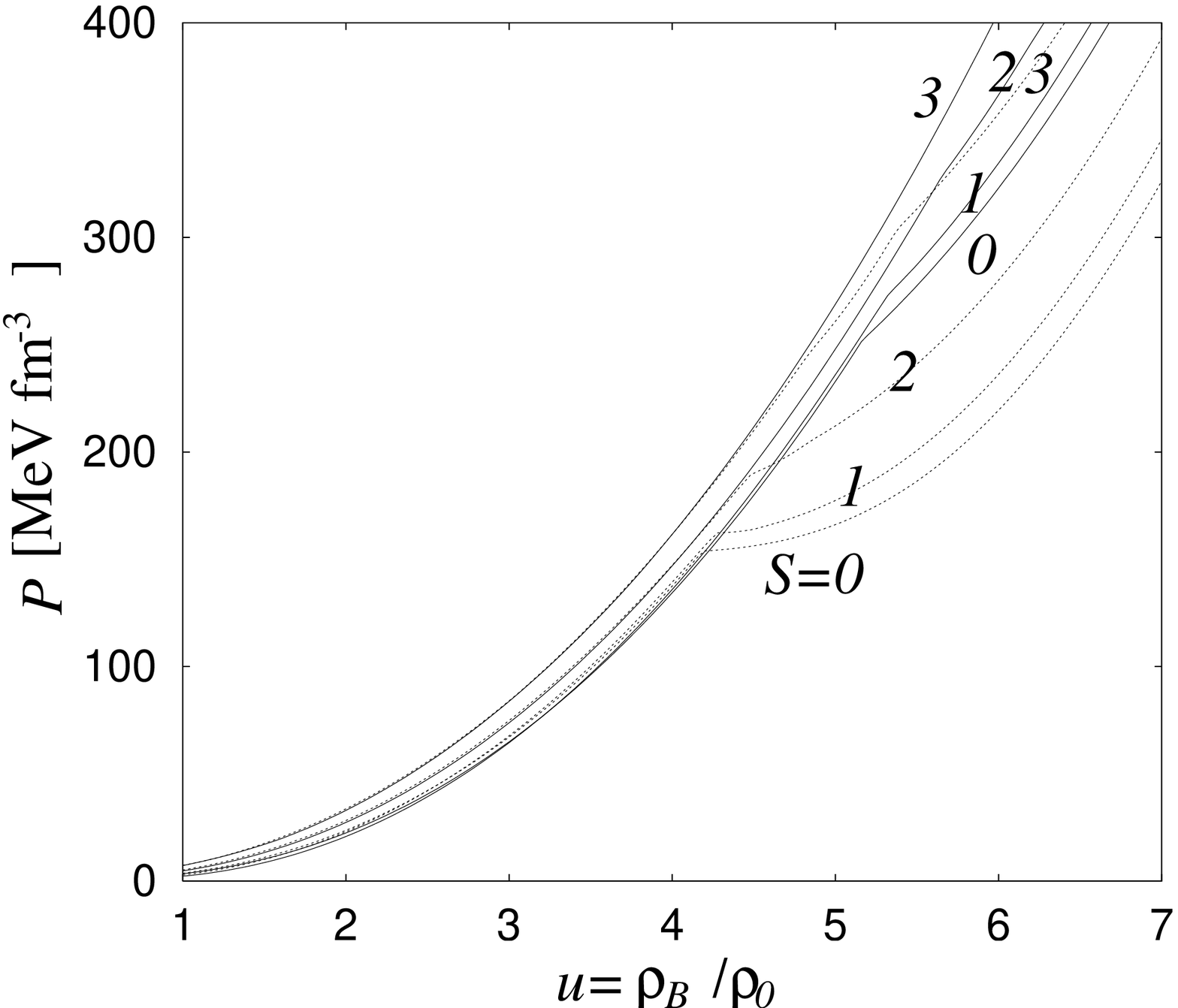}
 \end{minipage}%
 \hfill~%
 \begin{minipage}{0.49\textwidth}
  \epsfsize=0.99\textwidth
  \epsffile{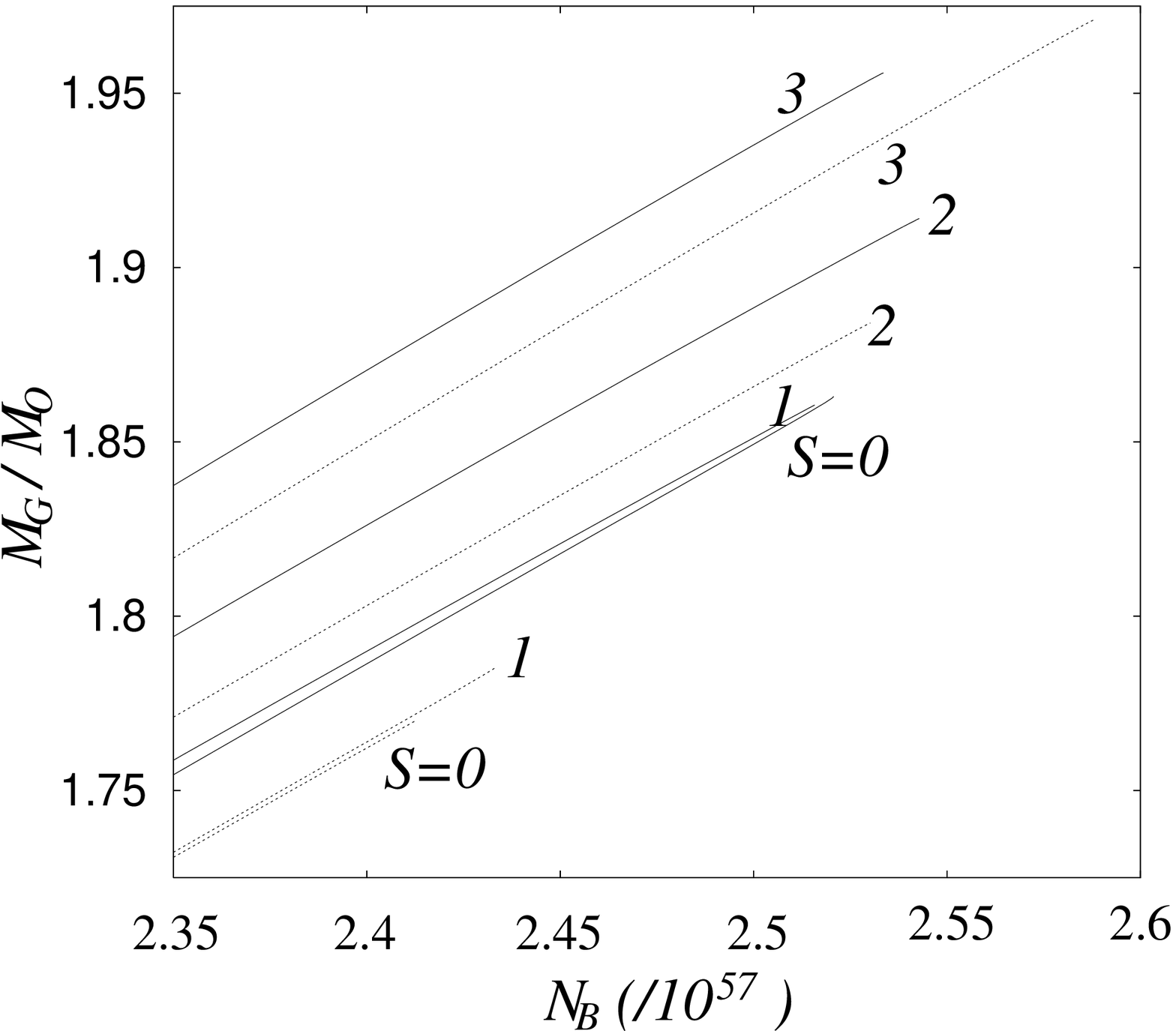}
 \end{minipage}%
 \caption{
The EOS [left panel]
and the relation between total baryon number
and gravitational mass
for stable PNS [right panel]
in the case of $\Sigma_{KN}=168$MeV.
Solid lines and dashed lines are
in neutrino-trapping ($Y_{le}=0.3$) and
neutrino-free case
at $S=0,1,2,3$.
}
 \label{fig:134}
\end{figure}

With these equations of state,
properties of the PNS can be studied as well.
The relations between the total baryon number
and the gravitational mass are shown
in the right panel in Fig.\ref{fig:134}.
We can also discuss the possibility of the
delayed collapse in the case of weak $KN$ coupling.
The behavior is very different from
that in the case of $\Sigma_{KN}=344$MeV in the left panel
in Fig.\ref{fig:NBMP}.
As entropy increases,
the terminal point in the neutrino-free case tends to move largely
to more baryon number and heavier gravitational mass
because their central densities remain around $6\rho_0$,
where the thermal effect in the EOS are remarkable in Fig.\ref{fig:134}.
On the other hand,
in the neutrino-trapping case,
the maximum total baryon number is
hardly changed in each configuration.

Then if we assume the same evolution in the above discussion:
a newly born PNS sitting on the curve in the
$S=2$ and neutrino-trapping case
evolves to the corresponding one on the curve
in the $S=2$ and neutrino-free case,
delayed collapse can occur in the deleptonization era.
Moreover,
it is also possible in the initial cooling era,
different from the case of $\Sigma_{KN}=344$MeV
(See the curves in the neutrino-free case in Fig.\ref{fig:134}).
Therefore we should conclude that
the thermal effect may cause the delayed collapse
as well as the neutrino-trapping effect
for the parameter $\Sigma_{KN}=168$MeV.

\section{Summary and Concluding Remarks}
\label{sect:sum}

In this paper,
using our new framework
based on chiral symmetry
to treat the kaon condensation at finite temperature,
we have discussed
the properties of kaon condensed matter and
its implication on astrophysics,
especially the delayed collapse.


When we discard any phase transition,
the delayed collapse is impossible
in deleptonization and initial cooling eras.
On the other hand,
when kaon condensation is realized,
the delayed collapse is possible.
We found that, in the case of $\Sigma_{KN}=344$MeV,
which is a most reasonable value for the $KN$ sigma term,
the neutrino-trapping effect gives rise to the delayed collapse
under no accretion of matter from the surroundings.
This point is a main difference
to the previous work about kaon condensation and
the delayed collapse\cite{Ponsprep}, in which
the meson-exchange model was used and it was concluded that
the thermal effect is a key object to the delayed collapse.

The difference may result
from the fact that, in the chiral model, the PNS
has very high central density,
compared with that in the meson-exchange model\cite{Ponsprep},
due to the stronger phase transition.
We also studied the case of weak limit of the $KN$ sigma term,
$\Sigma_{KN}=168$MeV,
which is a control parameter for the strength of 
the phase transition in the chiral model, to confirm our 
findings and see to what extent they are modified by
the parameter dependence.
We find that the neutrino-trapping effect
can cause the delayed collapse 
in the deleptonization era even in this case. 
However, we also find that the thermal effect becomes
important in this limit and the delayed collapse becomes 
possible by the thermal effect in some scenarios.
Some people may have impression that 
the Maxwell construction gives such high central density
in the chiral model.
However, we have rather high central density even
in the weak coupling limit, as is seen in Sec.\ref{sec:a3ms134}
where we no longer need the Maxwell nor Gibbs construction.
Hence we can say that the nonlinearity of 
the kaon field, which is an essential feature of the chiral model, 
gives rise to a stronger 
phase transition than in the meson exchange model. 
\footnote{The importance of the 
nonlinearity is also seen in the context of the discussion about the 
possibility of the Gibbs construction (see Appendix A).} 
Thus we conclude that 
the delayed collapse is possible
due to the occurrence of kaon condensation
in the core of the PNS and neutrino-trapping effect gives a main 
contribution in the chiral model.

In order to study the mechanism of the delayed collapse
and mass region of the PNS to collapse
in more detail,
we have to study the dynamical evolution
beyond the static discussion given here.
The accretion of a fall-back mass may
be an important ingredient\cite{WoosZampFry},
by which the delayed collapse may occur more easily.
Neutrino diffusion is another important ingredient
in that study:
we must treat the change of EOS,
neutrino opacity and heat capacity
in a consistent way
which come in the diffusion equation\cite{PonsH}\cite{PonsKsim}\cite{TT87s}.
Since the neutrino-nucleon scattering
or neutrino absorption by nucleons should be enhanced
in the kaon condensed phase\cite{MutoTYI}\cite{Reddy}
the neutrino-trapping era might last longer
if the heat capacity is little changed.

The heat capacity is related to the properties of matter
at finite temperature, which should be dominated
by the nucleon contribution\cite{Prakash}\cite{PonsH}\cite{Ponsprep}.
In the calculations with relativistic mean field theory(RMF),
the scalar interaction between kaon and nucleon
and that between nucleons may give contributions
through the modification of nucleon mass.
The effective mass of nucleons
affects the entropy or specific heat through
the equation for nucleon contribution to entropy per baryon ($S_N$),
\beq
  \frac{S_N}{\pi^2 T}
	= \frac{ x^{1/3}\sqrt{M_i^{*2}+(3\pi\rho_p)^{2/3}}
	  +(1-x)^{1/3}\sqrt{M_i^{*2}+(3\pi\rho_n)^{2/3}} }
	  { (3\pi^2\rho_B)^{2/3} }
  \label{eq:heat}
\eeq
with the low temperature approximation
in the relativistic treatment\cite{PonsH}\cite{Ponsprep}.
The nucleon effective mass is simply written as
$M_i^* = M - \Sigma_{Ki}(1-\cos\langle\theta\rangle)$, ($i=n, p$)
in our case,
and there is no contribution from the nucleon-nucleon interaction.
Then the ratio $M_i^*/M$ remains one in the normal phase
while it decreases to be 0.6-0.7 in the well-developed
kaon condensed phase.
In the realistic treatments of nuclear matter
it should have density dependence even in the normal matter,
so that it should be a smaller value than we have found;
e.g., it has been found 0.1-0.4
within the RMF\cite{Ponsprep}.
Furthermore we have treated nucleons
in a non-relativistic way,
so that the heat capacity is given only by their kinetic energies
without any effect by the condensate.
Then in the formula of entropy or specific heat
the bare mass comes in instead of the effective mass
and it may lead to the overestimation for specific heat.
At the same time
entropy may be overestimated as well;
we may underestimate the thermal effect in this paper.
In order to improve these points
resulting from the nucleon's effective mass,
we are planning to incorporate the RMF
besides the chiral model for the $KN$ interaction,
as an extension from
the previous work at zero temperature\cite{Fujii96}.


In this paper we studied the kaon condensation
and found
that the delayed collapse is possible
due to not only
the thermal effect but also
the neutrino-trapping effect
through the suppression of condensate.
It is very interesting to study
whether the delayed collapse occurs
in the deleptonization or initial cooling era
for other phase transitions, for example, hyperonic matter\cite{PonsH}
or quark matter.

\section{Acknowledgement}
We thank
T. Harada, K. Nakao, T. Takatsuka, T. Muto
and D.N. Voskressenski
for useful discussions and comments.

This work was supported in part by the Research Fellowships
of the Japan Society for the Promotion of Science
for Young Scientists
and by the Japanese Grant-in-Aid
for Scientific Research Fund of
the Ministry of Education, Science,
Sports and Culture(11640272).

\newpage
\section*{Appendix A}
In this paper we have applied the Maxwell construction (MC)
to get the EOS in equilibrium.
Strictly speaking,
we need to apply the Gibbs conditions (GC)
instead of using the MC\cite{gleGENE}
because there exist two chemical potentials:
baryon (neutron)
and charge (electron) chemical potentials.
Recently an attempt
appeared along this line to derive 
a realistic EOS\cite{gle}.
Then EOS should be smoothed due to
the appearance of the {\it mixed phase},
where normal matter
and kaon condensed matter coexist.
At the beginning of the mixed phase,
matter is expected to contain droplets
of, negatively charged, kaon-condensed matter
immersed in, positively charged, normal matter.
However some works suggest that
sometimes
the GC cannot be satisfied
\cite{Ponsprep}\cite{Heisel}.
Here,
we study the behavior of the thermodynamic potential
in each case of the Maxwell or Gibbs construction,
and show that the GC cannot
be satisfied in the chiral model
as already suggested by Pons et al.\cite{Ponsprep}.


Consider the mixed phase, where the bulk normal matter($N$)
and the bulk kaon condensed matter($K$) contact with each other
by a sharp boundary\footnote{
We discard any effect given by the boundary or the Coulomb energy.
}.
The MC demands the relations
\beq
  \mu_n^{N} = \mu_n^{K},\quad P^{N}=P^{K},
\eeq
on the other hand,
the GC can be written as
\beq
  \mu_n^{N} = \mu_n^{K},\quad
	\mu_e^{N} = \mu_e^{K},\quad
	P^{N}=P^{K},
  \label{eq:gibbs}
\eeq
between two phases.
Then we can see that the MC corresponds to
the Gibbs construction in the case of only one chemical potential.
First of all we figure out how the MC is applied in our study
by using thermodynamic potential.

In Fig.\ref{fig:omegaMC},
thermodynamic potential $\Omega_{total}/V$
is shown as a function of density or $\langle\theta\rangle$.
The dotted lines in both panels mean
the solutions under the equilibrium conditions;
chemical equilibrium Eq.(\ref{eq:chem}),
local charge neutrality Eq.(\ref{eq:cha}) and
equation of motion for kaon at momentum zero Eq.(\ref{eq:feq}).
Then the dotted lines show the EOS
through the relation $\Omega_{total}/V = -P$
in the left panel,
where normal matter with $\langle\theta\rangle=0$ exists
at low density $\rho_B<\rho_{crit}=3.1\rho_0$
and as density increases from $\rho_{crit}$,
thermodynamically unstable region ($\partial P/\partial \rho_B<0$)
appears up to $4.5\rho_0$
and the stable region again appears beyond that.
On the other hand,
the solid lines are calculated
with baryon chemical potential
fixed ($\mu_n=1032, 1082$ and $1132$MeV),
keeping the conditions for chemical equilibrium Eq.(\ref{eq:chem})
and local charge neutrality Eq.(\ref{eq:cha}).
In this case the cross points of the solid line with $\mu_n=1082$MeV
and the dotted line in the thermodynamically stable region
give the beginning and ending densities
of the equal-pressure region in Fig.\ref{fig:isoTuPP}.
Hence, if we use the MC
for the dotted line in the left panel,
we get the EOS for the $T=0$ and neutrino-free matter in Fig.\ref{fig:isoTuPP}.
In the right panel,
thermodynamic potential in the condensed phase is
given as a function of the order parameter $\langle\theta\rangle$.
The dotted line is given by connecting the extrema satisfying
$\partial\Omega_{total}/\partial\langle\theta\rangle=0$
in solid lines, which corresponds to the dotted line
in the left panel.
It is to be noted that the thermodynamically unstable region
in the left panel exactly corresponds to the left region
to the maximum of the dotted line in the right panel
where $\partial^2\Omega_{total}/\partial\langle\theta\rangle^2<0$.
The minimum in each solid line means the ground state;
in the case of low $\mu_n$ the minimum point stays
at $\langle\theta\rangle=0$,
which means the normal phase is favored.
Afterwards two minima appear at $\mu_n=1082$MeV
with equal pressure, corresponding
the {\it coexisting phase}\footnote{
In this section, we call the mixed phase in the case
using the MC as the ``coexisting phase''
and in the case imposing the GC, the ``mixed phase''
to avoid confusion.}
in the context of the MC.
Above this value of baryon chemical potential,
the kaon condensed phase is always favored.
\begin{figure}[htb]
 \vspace{2mm}
 \begin{minipage}{0.48\textwidth}
  \epsfsize=0.99\textwidth
  \epsffile{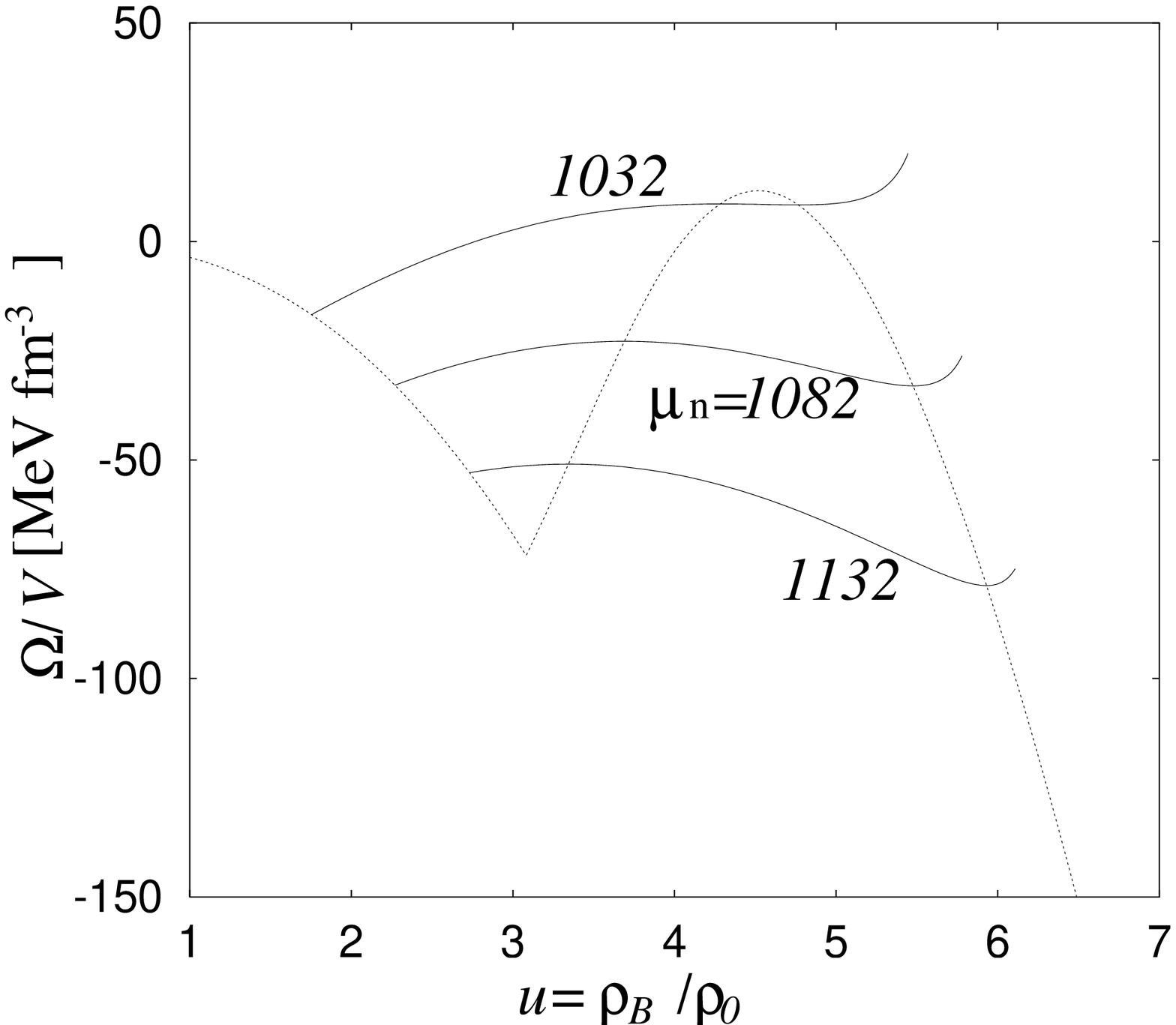}
 \end{minipage}%
 \hfill~%
 \begin{minipage}{0.48\textwidth}
  \epsfsize=0.99\textwidth
  \epsffile{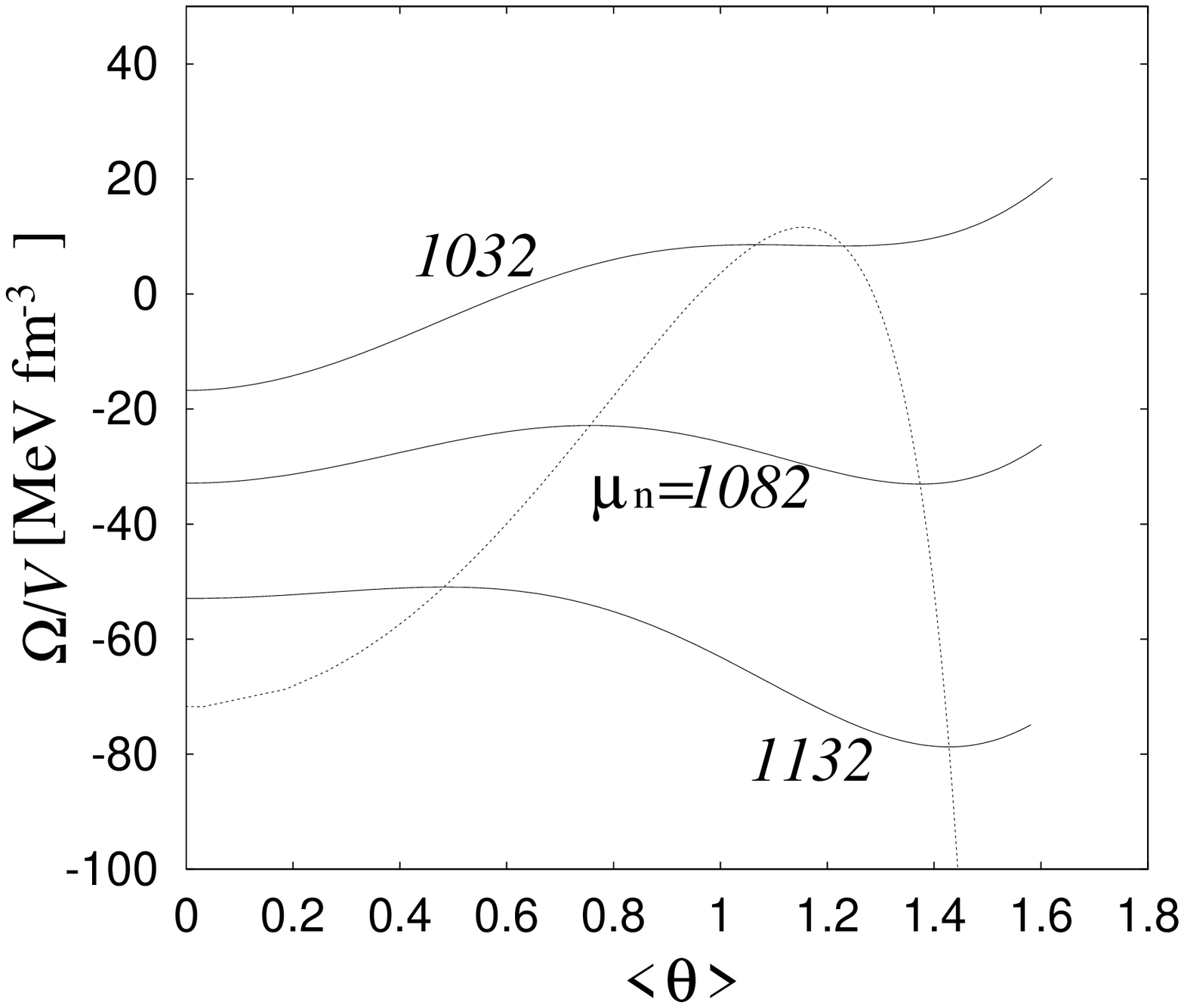}
 \end{minipage}%
 \caption{The behavior of thermodynamic potential
is shown as a function of density [left panel]
and $\langle\theta\rangle$ [right panel].
$\mu_n$ is fixed and chemical equilibrium and
local charge neutrality are achieved in solid lines.
Dotted lines mean the solutions of equilibrium conditions;
chemical equilibrium, local charge neutrality and
equation of motion for kaon.
}
 \label{fig:omegaMC}
\end{figure}

Using the MC,
typical characteristics are found
for the structure of neutron stars as follows;
1: Gravitationally unstable region
($\partial M_G/\partial \rho_c < 0$) appears.
2: There is no {\it coexisting phase} in neutron stars\footnote{
As already we noted in Sect.\ref{sec:eos},
this comment is correct only when we construct
the neutron stars with zero entropy.
Finite entropy allows the existence of the {\it coexisting phase}.
}.
3: Equal-pressure region appears in the EOS
and it leads to the density gap in the structure of a neutron star.

These features will disappear
when we apply the GC,
which is a correct treatment
when there exist more than one chemical potential\cite{gle}.
In order to discuss whether
the GC can be satisfied or not,
by the similar manner as in the case of the MC,
the behavior of thermodynamic potential $\Omega_{total}/V$
is shown in Fig.\ref{fig:omegaGC}
with two chemical potentials fixed
at $\mu_n = 1082$MeV and $\mu_e = -100, -50,$$\cdots 150$MeV,
keeping chemical equilibrium.
If the GC can be satisfied,
the existence of two minima is expected.
However we find only single minimum at $\langle\theta\rangle=0$.
Then we can conclude the GC cannot be satisfied
in the chiral model\cite{Ponsprep}.
On the other hand, in the meson-exchange model,
two minima exist and the GC can be applied
except the strong phase transition\cite{Ponsprep}.
The difference between two models
results from the nonlinearity of kaon field
in the chiral model
and we can see that
even if we use the weak $KN$ sigma term,
the GC cannot be satisfied in the chiral model\footnote{
Using the chiral model and the
``potential energy'' contribution Eq.(\ref{eq:npot})
for $NN$ interaction
in non-relativistic approximation,
the Gibbs conditions cannot be satisfied,
even in the case of the weak $KN$ sigma term,
for example, in the case of $\Sigma_{KN}=180$MeV
where appears very narrow thermodynamically unstable region.
It may also be interesting to study
the case of relativistic treatment for $NN$ interaction
besides the chiral model for $KN$ interaction.
}.
\begin{figure}[ht]
 \vspace{2mm}
  \epsfsize=0.5\textwidth
  \epsffile{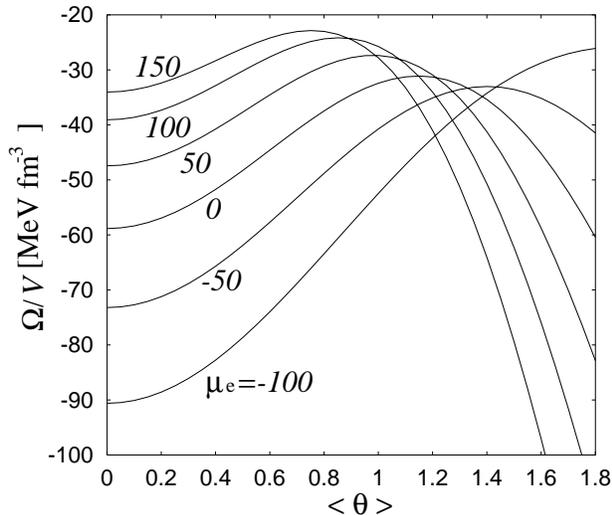}
 \caption{The behavior of thermodynamic potential
is shown as a function of $\langle\theta\rangle$.
Two chemical potentials are fixed
and chemical equilibrium is achieved in each line.
}
 \label{fig:omegaGC}
\end{figure}

It is to be noted that we have discussed
the thermodynamic potential
in the ideal situation
where the surface and Coulomb effects are discarded.
However, this type of argument may miss as essential point
for the existence of the mixed phase.
When we consider a realistic situation for the mixed phase,
there is no clear boundary between two phases;
the density of each constituent and the chiral angle
should be smoothly changed there.
So the naive GC Eq.(\ref{eq:gibbs})
cannot be applied to this situation.
Instead, we must carefully determine the configuration
of the mixed phase by allowing the chemical potentials
to depend on the spatial coordinates and taking into
account the screening effect of the electric potential.
Then we can compare the energy of the mixed phase
with the normal phase to see whether the mixed phase
can exist.
Recently there have been appeared some works by taking
into account the surface and Coulomb effects
along this line\cite{Reddy}\cite{droplet}.
In this treatment, we think, there is some possibility
that the chiral model may give the mixed phase as well.
This interesting issue is left for a future study.

\newpage

\end{document}